\documentclass[11pt]{article}
\usepackage{jheppub}
\usepackage{amsmath}
\usepackage{verbatim}   
\usepackage{subfigure}  
\usepackage{amsfonts}
\usepackage{amssymb}
\usepackage{graphicx}
\usepackage{epsfig}
\usepackage{url}
\usepackage{epigraph}

\newcommand{ \EW }{\ensuremath{SU(3)_{\rm EW} }}
\newcommand{ \SMLY }{\ensuremath{SU(2)_{\rm L} \times U(1)_{\rm Y} }}
\newcommand{ \SUC }{\ensuremath{SU(3)_{\rm C} }}
\newcommand{ \diag }{\ensuremath{\rm{diag} }}
\newcommand{\TeV}{\,{\rm TeV}}
\newcommand{\GeV}{\,{\rm GeV}}

\def\beq{\begin{equation}}
\def\eeq{\end{equation}}
\def\bea{\begin{eqnarray}}
\def\eea{\end{eqnarray}}
\def\bitem{\begin{itemize}}
\def\eitem{\end{itemize}}
\newcommand{\bec}{\begin{center}}
\newcommand{\eec}{\end{center}}
\newcommand{\ba}{\begin{array}}
\newcommand{\ea}{\end{array}}

\title{Unified Maximally Natural Supersymmetry}

\abstract{Maximally Natural Supersymmetry,  an unusual weak-scale supersymmetric extension of the Standard Model 
based upon the inherently higher-dimensional mechanism of Scherk-Schwarz supersymmetry breaking (SSSB),
possesses remarkably good fine tuning given present LHC limits.  Here we construct a version with precision $\SMLY$
unification: $\sin^2 \theta_W(M_Z)\simeq 0.231$ is predicted to $\pm 2\%$ by unifying $\SMLY$ into a 5D $\EW$ theory at 
a Kaluza-Klein scale of $1/R_5 \sim 4.4 \TeV$, where SSSB is simultaneously realised.   Full unification with $\SUC$
is accommodated by extending the 5D theory to a $N=4$ supersymmetric $SU(6)$ gauge theory on a 6D rectangular orbifold at $1/R_6 \sim 40 \TeV$.
TeV-scale states beyond the SM include exotic charged fermions implied by $\EW$ with masses lighter than $\sim 1.2 \TeV$, and squarks in the mass range
$1.4 \TeV - 2.3 \TeV$, providing distinct signatures and discovery opportunities for LHC run II.}

\author[a]{Junwu Huang,}
\emailAdd{curlyh@stanford.edu}
\author[b]{John March-Russell}
\emailAdd{jmr@thphys.ox.ac.uk}

\affiliation[a]{Stanford Institute for Theoretical Physics, Department of Physics, Stanford University,\\
Stanford, CA 94305, USA}
\affiliation[b]{Rudolf Peierls Centre for Theoretical Physics, University of Oxford, 1 Keble Road,\\
Oxford, OX1 3NP, UK}

\begin{document}

\maketitle

\pagebreak

\section{Introduction}

Recently~\cite{Dimopoulos:2014aua,Garcia:2014lfa,Garcia:2015sfa} there has been renewed interest in supersymmetric (SUSY) models that resolve the little hierarchy problem of the Standard Model~\cite{Gherghetta:2012gb,Arvanitaki:2013yja,Feng:2013pwa,Fan:2014txa,Gherghetta:2014xea,Hardy:2013ywa} by utilizing the remarkable properties of the Scherk-Schwarz supersymmetry breaking (SSSB) mechanism~\cite{Scherk:1978ta,Scherk:1979zr}.  The SSSB mechanism, previously applied to weak-scale SUSY model building by a variety of authors~\cite{Antoniadis:1998sd,Delgado:1998qr,Pomarol:1998sd,Delgado:2001si,Barbieri:2003kn,Barbieri:2002sw,Barbieri:2002uk,Barbieri:2000vh,Diego:2005mu,Diego:2006py,Gersdorff:2007kk,Bhattacharyya:2012ct,Quiros:2003gg,Delgado:2001xr}, breaks SUSY by boundary conditions (bc's) in one (or more) extra dimensions.  The new implementation of this idea, so-called Maximally Natural Supersymmetry (MNSUSY)~\cite{Dimopoulos:2014aua,Garcia:2014lfa,Garcia:2015sfa} is consistent with LHC run I constraints on superpartners, rare process and flavor constraints, and achieves the observed higgs mass, while maintaining a low fine tuning of $\sim 20\%$.  This improvement
in tuning compared to MSSM-like SUSY models mostly follows from the fact that SSSB locally maintains unbroken SUSY at each point of the extra dimension, with SUSY only being broken non-locally with respect to the extra dimension.   This non-locality of SUSY breaking then protects the higgs boson from getting large log-enhanced loop corrections to its soft mass squared~\cite{Antoniadis:1997zg,Barbieri:2001dm,Contino:2001gz,Delgado:2001ex}.  For maximal SSSB of an underlying 5D SUSY theory, the case utilized in the MNSUSY construction, the extra dimension may equivalently be thought of as an $S^1/(Z_2\times Z_2')$ orbifold with boundary fixed points, and we will use this description in the rest of this work.\footnote{{\it Gauge} symmetry breaking by orbifold bcs, also known as the Hosotani mechanism~\cite{Hosotani:1983xw,Hosotani:1983vn,Hosotani:1988bm}, is an important ingredient in constructing simple, realistic grand unified theories (GUTs), as it gives an elegant solution to the doublet triplet splitting problem~\cite{Kawamura:2000ev,Altarelli:2001qj,Hall:2001pg} while maintaining precision SUSY gauge-coupling unification~\cite{Hall:2001pg,Hebecker:2001wq,Hebecker:2001jb,Hall:2001xb}.}

However, the SSSB-based models studied so far, including MNSUSY, give up one of the most attractive successes of MSSM-like theories, namely precision gauge coupling unification and the associated prediction of the weak mixing angle $\sin^2 \theta_W(M_Z)$~\cite{Dimopoulos:1981zb,Dimopoulos:1981yj}.  
In conventional SUSY theories, the unification of couplings is realised with differential logarithmic running of Standard Model gauge couplings up to a high scale, $\sim 2\times 10^{16}\GeV$, but this is not possible with extra dimensions at the TeV scale and an associated cutoff much below $10^{16}\GeV$.  

On the other hand, the relative closeness between the observed $\sin^2 \theta_W^{\overline{\rm MS}}({M_Z}) \approx  0.23116\pm 0.00012$, and $1/4$ motivate consideration of models where $\SMLY$ is unified {\it at low scales} into an $\EW$ gauge group~\cite{Dimopoulos:2002jf,Dimopoulos:2002mv,Li:2002pb,Hall:2002rk} where $\sin^2 \theta_W =1/4$ is the unified prediction before logarithmic radiative corrections.  This, of course, is only a partial unification, but in principle $\EW$ can be unified with $\SUC$ into a simple group in 6D~\cite{Hall:2002qw,Jiang:2002at,Hall:2001zb,Hartanto:2005jr}. In this work, we construct a unified 6D model where the effective theory at low energies is MNSUSY (with some
specified exotic states), and we study its predictions and low-energy phenomenology.

\begin{figure}[htb]
\centering
\begin{minipage}[t]{0.50\textwidth}
 \includegraphics[width=\textwidth]{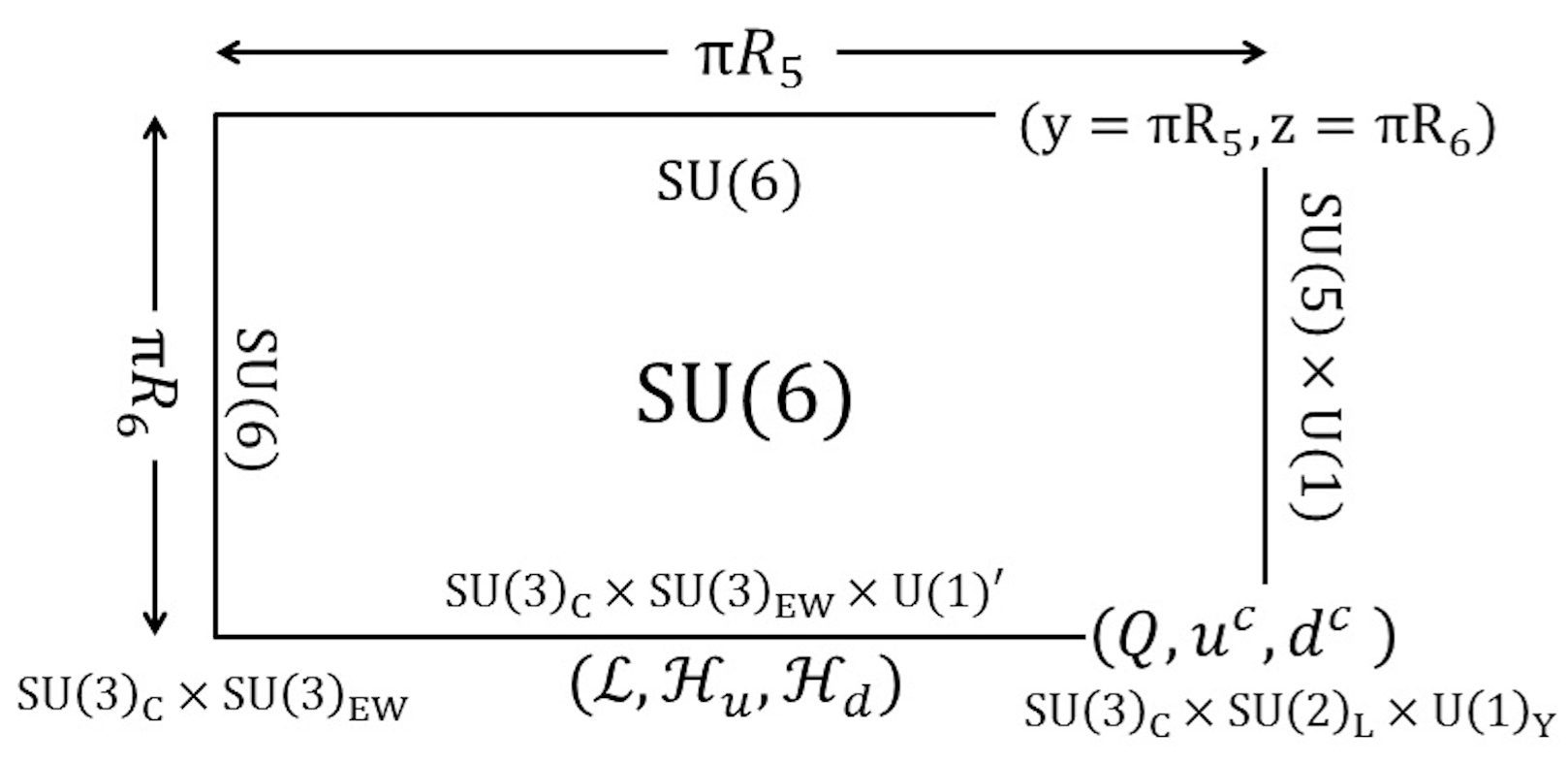}
\end{minipage}
\begin{minipage}[t]{0.45\textwidth}
\centering
\includegraphics[width=\textwidth]{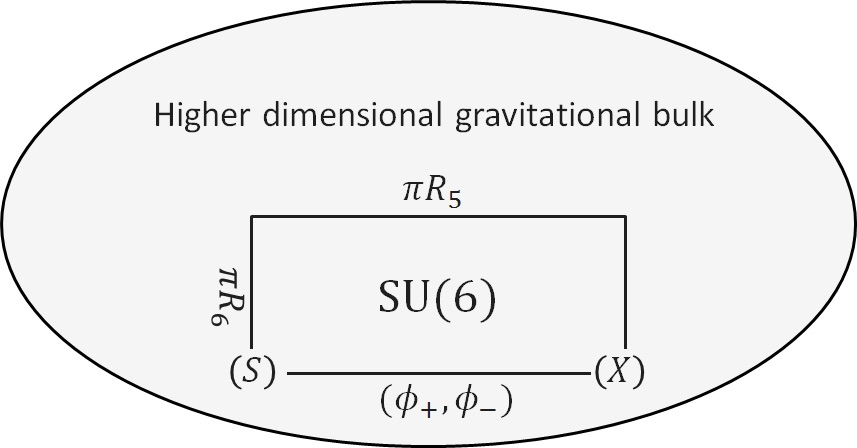}
\end{minipage}
\caption{{\it Left panel:} Schematic illustration of the symmetry structure of the minimal $SU(6)$ unified model. In this plot,
each line is a (4+1)-dimensional orbifold line and each corner is a (3+1)-dimensional orbifold fixed point in the two extra
dimensions $(y,z)$ of the total 6D space. The gauge symmetry on each brane is as indicated, or the unbroken $SU(6)$ otherwise. The bulk $N=4$ supersymmetric $SU(6)$ gauge symmetry is broken by orbifold bcs. The gauge symmetry on the bottom line ($z=0$ brane) is $\SUC \times \EW \times U(1)'$. The leptons ($\mathcal{L} = (L, e^c)$) and higgses ($\mathcal{H}_{u,d} = (H_{u,d},S_{u,d})$) are simultaneously 5D SUSY hyper-multiplets and triplets respecting the $\EW$ gauge symmetry. The quark chiral multiplets $Q,u^c,d^c$ do not respect the $\EW$ gauge symmetry and therefore live on the $y=\pi R_5,z=0$ brane where the gauge symmetry is broken to the SM gauge group.  Different, incompatible $N=1$ SUSYs are realized on the $y=\pi R_5,z=0$ and $y=0,z=0$ branes, resulting in non-local SUSY-breaking on the 5 dimensional interval. {\it Right panel: } The embedding of the 6D theory into a higher dimensional gravitational bulk with Standard Model singlets indicated. These singlet states are required to break the $U(1)'$ gauge symmetry and generate masses for the light modes.}\label{Fig::su6}
\end{figure}

A schematic illustration of our 6D model is shown in Figure~\ref{Fig::su6} (the possible further embedding in a yet higher-dimensional gravitational bulk is shown in the right panel of Figure~\ref{Fig::su6}).  A $N=4$ supersymmetric\footnote{Throughout this work the size of the supersymmetry algebra is expressed in terms of the corresponding supersymmetry algebra in four dimensions.} $SU(6)$ gauge theory is compactified on a $S^1/(Z_2\times Z_2') \times S^1/(Z_2 \times Z_2')$ orbifold with symmetry structure as shown in Figure~\ref{Fig::su6}~\cite{Hall:2002qw,Jiang:2002at}. The leptons and higgses, now extended to full $\EW$ triplets, ($\mathcal{L} = \{L, e^c\}$, $\mathcal{H}_{u,d}
= \{ H_{u,d},S_{u,d}\} $), are localized on the (4+1)-dimensional brane at $z=0$ preserving $\SUC \times \EW \times U(1)'$ and $N=2$ SUSY.  In distinction, the quark super-multiplets $Q, u^c, d^c$ cannot be embedded in $\EW$, and must be localized on the (3+1)-dimensional orbifold
fixed point at $y= \pi R_5, z=0$ where just the SM gauge group and $N=1$ SUSY are good symmetries.

The benchmark size of the fifth dimension is chosen to be $1/R_5\sim 4.4 \TeV$ to be compatible with experiment and at the same time minimize the fine-tuning, while $R_6$ is only constrained by the unitarity bound on gauge boson scattering in higher-dimensional
gauge theories, as we discuss in Section~\ref{Sec::unitarity}.  In the benchmark model, we have an effective GUT scale
$M_{\rm GUT} \sim 10 (1/\pi R_6) \sim 90 (1/\pi R_5)\sim 100 \TeV$.  (Throughout the paper, we work in the limit $M_{\rm GUT},1/R_6 \gg 1/R_5$.)

\begin{figure}[htb]
\centering
\includegraphics[width=0.9\columnwidth]{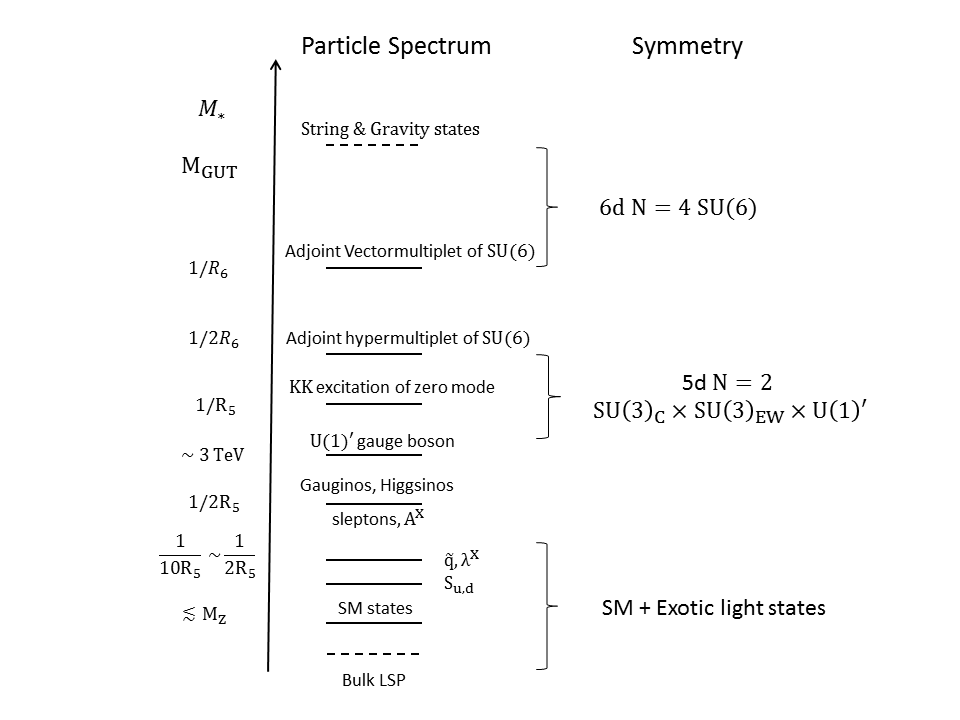}
\caption{Schematic spectrum of states that are of experimental interest, and states that are required so as to get $SU(6)$ unification. The symmetry of the two stage unification model at various energy scales is shown on the right of the plot. The energy axis is not to scale.}\label{Fig::spectrum}
\end{figure}

$\SMLY$ unifies into $\EW$ due to differential logarithmic running from $M_Z$ to $M_{\rm GUT}$~\cite{Dimopoulos:2002jf,Dimopoulos:2002mv,Li:2002pb,Hall:2002rk}. The beta functions can be calculated in the full energy range and calculable $\EW$ breaking threshold corrections at $1/R_5$ can be obtained by the method in~\cite{Contino:2001si}.
On the other hand, the non-calculable $\EW$ breaking threshold corrections are estimated using 
the strong coupling analysis~\cite{Chacko:1999hg}.  The scale where $\SMLY$ unifies into $\EW$ depends on only physics below or around $1/R_5$, namely, the squark masses, the heavy higgs masses and masses of gauginos and singlet higgs zero modes.  A detailed discussion of this $\EW$ unification is contained in Section~\ref{Sec::SU3}.

Moving on to the more ambitious objective of a full unification of $\EW\times\SUC$ into a single, higher dimensional
GUT structure, we utilize recent knowledge of extended SUSY theories in higher dimensions~\cite{Seiberg:1996bd,Intriligator:1997pq,Seiberg:1996qx,Danielsson:1997kt}. In particular, by studying the properties of the pre-potential, it has been shown that higher-dimensional operators involving gauge fields at two derivative order are highly restricted by $N=2$ SUSY and gauge symmetry, and that power-law threshold corrections can be, in some cases, exactly calculated~\cite{Hebecker:2002vm,Hebecker:2004xx}. These analyses lay the foundation of our study of precision unification by power-law threshold corrections.

Specifically, we argue that $\SUC \times \EW$ unifies into $SU(6)$ on a 6D orbifold due to differential logarithmic running from $M_Z$ to $M_{\rm GUT}$ and, dominantly, calculable power (linear) threshold corrections~\cite{Hebecker:2002vm,Hebecker:2004xx},the differential linear threshold corrections being due to the non-$SU(6)$ symmetric matter content on the $z=0$ brane. The scale where $\SUC \times \EW$ unifies into $SU(6)$ depends on dominantly the size, $\pi R_5$, of the $z=0$ brane together with its particle content.  A detailed discussion of this $SU(6)$ unification is in Section~\ref{Sec::SU6}.

The $SU(6)$ unifying group also contains, after breaking by orbifold bcs, a surviving $U(1)'$ gauge symmetry in addition
to the SM.  In fact such a symmetry was posited in the original MNSUSY model~\cite{Dimopoulos:2014aua,Garcia:2015sfa} so as to raise the
physical higgs mass to its observed value. $U(1)'$ is broken by vacuum expectation values (vevs) of the 5D fields, $\phi_{\pm}$, and a contribution from the non-decoupling D-term~\cite{Maloney:2004rc} of the broken $U(1)'$ lifts $m_h$ to $125 \GeV$.  A detail account of this dynamics in our context can be found in Section~\ref{Sec::U1prime}.

Finally, in Section~5 we discuss the collider phenomenology of our unified model, focussing in particular
on the aspects that differ from the basic MNSUSY model, and also on issues directly related to unification.  Interestingly we find that the broken $U(1)'$, together with $U(1)_R$ symmetry and locality forbid the dangerous proton decay and neutron anti-neutron oscillation operators (Section~\ref{sec::protondecay}).


\section{$\EW$ unification}\label{Sec::SU3}

In this section, we discuss how $\EW$ unification can be realized with one TeV-sized dimension in a model that 
simultaneously realizes the MNSUSY scenario at the TeV-scale.  In later sections we will go on to show how this 5D $N=2$ SUSY
theory can be extended to a full $SU(6)$ $N=4$ SUSY unification in 6D.  Our 5D model is based upon the non-supersymmetric Dimopoulos-Kaplan $\EW$ model~\cite{Dimopoulos:2002jf,Dimopoulos:2002mv,Hall:2002rk,Li:2002pb}, but we extend their construction to realize SSSB, and perform a precision calculation of the prediction for $\sin^2\theta_W$. 

\subsection{Minimal matter content and orbifold boundary conditions}

Consider a 5D $N=2$ SUSY $\EW$ gauge theory compactified on a $S^1/Z_2 \times Z'_2$ orbifold line segment (where $y\in [0,\pi R_5]$) with bc's on the $\EW$ gauge fields $A_{\mu}\equiv A_{\mu}^a T^a$ determined by the orbifold equivalences
\begin{align}
A_{\mu} (x^{\mu},y) &= A_{\mu} (x^{\mu},- y) ~~{\rm and}\nonumber\\ 
 A_{\mu} (x^{\mu},y) &= Z' A_{\mu} (x^{\mu}, - y + 2 \pi R_5) Z'^{-1} 
\end{align}
where $Z' = \diag(1,1,-1)$ acts on $\EW$ indices, and the action on the $A_5$ components has an extra overall minus sign in both
relations.   We need to extend this action to the full field content of the theory.  As minimal 5D bulk SUSY corresponds, from a 4D perspective, to $N=2$ SUSY. Let us
first recall how $N =2$ SUSY is written in terms of 4D $N=1$ supermultiplets:  The vector supermultiplet of $N =2$
contains a $N=1$ vector supermultiplet, $V$, and a chiral supermultiplet, $\tilde{\Sigma}$, both adjoint 
representation fields.  In terms of on-shell degrees of freedom $V= (A_\mu, \lambda)$ and $\Sigma = (\tilde{\Sigma}, \psi)$, where
$\lambda$ and $\psi$ are Weyl fermions (the complex scalar field $\tilde{\Sigma}$ contains $A_5$).  Similarly, the $N=2$ hypermultiplet contains two 4D $N=1$ chiral supermultiplets with conjugate gauge quantum numbers and R-charge.

These bc's under $(Z_2,Z_2')$ are extended to the component fields of the full vector multiplet
\begin{align}
 &A_\mu : \begin{pmatrix}
(+,+) & (+,+) & (+,-)   \\
(+,+) & (+,+) & (+,-)  \\
(+,-) & (+,-) & (+,+)
\end{pmatrix}
 &\tilde{\Sigma} : 
\begin{pmatrix}
(-,-) & (-,-) & (-,+) \\
 (-,-) & (-,-) & (-,+) \\
(-,+) & (-,+) & (-,-) 
\end{pmatrix} \nonumber\\
 &\lambda : \begin{pmatrix}
(+,-) & (+,-) & (+,+)   \\
(+,-) & (+,-) & (+,+)  \\
(+,+) & (+,+) & (+,-)
\end{pmatrix}
 &\psi : 
\begin{pmatrix}
(-,+) & (-,+) & (-,-) \\
 (-,+) & (-,+) & (-,-) \\
(-,-) & (-,-) & (-,+) 
\end{pmatrix} ,
\end{align}
where we have specified the bc's that apply to the various $SU(3)$ components, with $+$ and $-$ corresponding to Neumann and Dirichlet respectively.  This action breaks $\EW$ down to $\SMLY$ on the 4D orbifold brane at $y=\pi R_5$.  
Moreover, the bc's of gauginos and scalars are such that {\it two different} $N=1$ SUSYs are preserved locally on each boundary, while the incompatibility of the two $N=1$ SUSYs breaks SUSY completely and non-locally by (maximal) SSSB.
The $(+,+)$ zero modes in this sector of the theory include the $\SMLY$ gauge bosons $A_{\mu}^{\rm EW}$ and the gauginos $\lambda^{\rm{X}}$.  The KK tower for the vector multiplet contains the modes with $(+,+)$ and $(-,-)$ bc's at mass $(n+1)/R_5$, and the modes with $(+,-)$ or $(-,+)$ bc's at mass $(n+1/2)/R_5$.

Since the quark chiral supermultiplets, $Q, u_R$ and $d_R$, have hypercharge assignments that do not descend from $SU(3)$ representations, they have to be localized on the $y=\pi R_5$ brane.  On the other hand, the higgs $H_{u,d}$ and leptons $L,e_R$ (of all 3 generations) can, in principle, be either bulk or brane localized fields consistent with $\EW$.  The analysis of the MNSUSY model in~\cite{Dimopoulos:2014aua} shows that if the 
higgs is a bulk field then a low-fine-tuning solution to the little hierarchy problem is possible by virtue of the SSSB bc's.  Thus we extend the higgs multiplets to full $\EW$ triplets by the addition of fields $S_{u,d}$, giving triplets $\mathcal{H}_{u,d} \equiv (H_{u,d},S_{u,d})$.  In this way
the higgs fields consistently transform under the 5D bulk $\EW$ gauge symmetry.  Similarly, we choose here to place the leptons also in the bulk as triplets $\mathcal{L} \equiv (L,e_R)$ -- this allows suitable lepton Yukawas to be generated (See equation~\ref{Eq:yukawa}).  The orbifold bc's for the scalar and fermion components of the resulting higgs and lepton hypermultiplets are
\begin{align}
 &\mathcal{H}_{u,d} : \begin{pmatrix}
 (+,+) \cr (+,+) \cr (+,+)
 \end{pmatrix}, ~~
&\mathcal{H}_{u,d}^c : \begin{pmatrix}
 (-,-) \cr (-,-) \cr (-,-)
 \end{pmatrix}, ~~
&\tilde{\mathcal{H}}_{u,d} : \begin{pmatrix}
 (+,-) \cr (+,-) \cr (+,-)
 \end{pmatrix}, ~~
&\tilde{\mathcal{H}}_{u,d}^c : \begin{pmatrix}
 (-,+) \cr (-,+) \cr (-,+)
 \end{pmatrix}.\nonumber\\
 & \begin{pmatrix}
 L \cr e_R
 \end{pmatrix} : \begin{pmatrix}
 (+,+) \cr (+,+) \cr (+,+)
 \end{pmatrix}, ~~
 &\begin{pmatrix}
 L^c \cr e_R^c
 \end{pmatrix} : \begin{pmatrix}
 (-,-) \cr (-,-) \cr (-,-)
 \end{pmatrix}, ~~
&\begin{pmatrix}
 \tilde{L} \cr \tilde{e}_R
 \end{pmatrix} : \begin{pmatrix}
 (+,-) \cr (+,-) \cr (+,-)
 \end{pmatrix}, ~~
&\begin{pmatrix}
 \tilde{L}^c \cr \tilde{e}_R^c
 \end{pmatrix} : \begin{pmatrix}
 (-,+) \cr (-,+) \cr (-,+)
 \end{pmatrix}.
\end{align}
With the above set of fields and bc's, the low energy effective theory well below the compactification scale $1/R_5$ includes all fields in the SM charged under the SM electroweak symmetry, the superpartners of the brane localized quark fields, a vector pair of zero
mode gauginos $\lambda^{\rm{X}}$ with quantum number $(2,\pm 3/2)$ under the $\SMLY$ symmetry
and a vector pair of scalars ${S}_{u,d}$ with quantum numbers $(1,\pm 1)$.

\subsection{Differential logarithmic running in SUSY orbifold theories}\label{Sec::difflog}

Precision unification into $\EW$ involves the relative logarithmic running of $g_1$ and $g_2$. We will briefly outline the argument~\cite{Contino:2001si} that shows that this relative running of the gauge couplings in a 5D SUSY orbifold theory can be calculated to high precision.\footnote{Although the effect of any brane localized charged matter is not included in the discussion below, it can be accounted by purely standard 4D calculations, as we will do when we present final results.} 

The first step is to write down all possible bulk and brane localized counterterms. The counterterm corresponding to the renormalization of the (dimensionful) 5D bulk gauge coupling has dimension $[mass]^1$ and therefore can, in principle, get both a linear and a logarithmic dependence on the UV cut-off, $\Lambda$, of the form $\Lambda + m \log \Lambda$, where $m$ is a mass parameter in the bulk Lagrangian.  In our case the linear piece is {\it universal} since the 5D bulk is $\EW$ symmetric while the logarithmic piece vanishes due to the absence of bulk mass parameters in our theory.  However, there is still coupling running above the compactification scale $1/R_5$ due to the existence of the orbifold fixed points where counterterms in the form of 
\begin{equation}
\mathcal{L}_{ct} = \int {\rm d}^4 x {\rm d} y \  F_{\mu \nu} F^{\mu \nu} \left[ a ~\delta(y) + b ~\delta (y - \pi R_5) \right]
\label{Eq:Lct5D}
\end{equation}
can be written down for a $S^1/Z_2 \times Z_2'$ orbifold, where $a$ and $b$ are constants that cancel the UV divergences of the orbifold gauge theory.  These counterterms in 5D can be Fourier decomposed in terms of the 4D KK gauge boson excitations
\begin{align}
\mathcal{L}_{ct} = \int {\rm d}^4 x & \left\{ (a+b)\left[ \sum_{n=0}^{\infty} c_{2n}^2 F^{(2n)}_{\mu \nu} F^{ (2n) \mu \nu} + \sum_{n=0,k=1}^{\infty} 2 c_{2n} c_{2n+4k} F^{(2n)}_{\mu \nu} F^{ (2n+ 4 k) \mu \nu} \right] \right. \nonumber\\
& \left.+ 2 (a-b) \sum_{n=0,k=1}^{\infty} c_{2n} c_{2n+4k-2} F^{(2n)}_{\mu \nu} F^{ (2n+ 4 k-2) \mu \nu} \right\} .
\label{Eq:Lct4D}
\end{align}

Next, from examination of the 5D KK momenta in loop diagrams (including the relevant KK mode identifications due to the orbifold bc's), it follows that the exchange of odd KK modes only produces transitions $2n \rightarrow 2n + 4k - 2$ while the exchange of even KK modes only produces transitions $2n \rightarrow 2n + 4k$.  Comparing with Eq.(\ref{Eq:Lct4D}) then implies that loop diagrams
involving odd KK modes lead to $a=-b$, while even KK modes in loops lead to $a=b$.  Since the gauge kinetic term of the {\it zero mode} is not renormalized when $a= -b$, the odd KK modes do not affect the running of the effective 4D gauge coupling.

The beta function of the whole KK tower compared with that arising from the corresponding zero mode can then be calculated and understood in the following manner~\cite{Contino:2001si,Hall:2001pg,Hall:2001xb,Barbieri:2001dm}. The absence of logarithmic running in the absence of the orbifold fixed points implies that the sum of the beta functions due to the exchange of the zero mode and the KK modes altogether is zero if the bulk is an $S^1$.  Moreover, in the case of $S^1$, KK modes with masses equal to the even modes and the odd modes with respect to the $S^1/Z_2 \times Z_2'$ orbifold have the {\it same} contribution to the beta function coefficient. Therefore, on $S^1$ both the even and odd modes contribute $-1/2$ to the gauge coupling running compared with the corresponding zero mode. Since, as we previously argued, in a $S^1/Z_2 \times Z_2'$ orbifold theory, the exchange of the odd modes does not contribute to the logarithmic running of gauge coupling, we finally get that {\it the logarithmic running of gauge coupling due to bulk states slows down above the compactification scale to half the original rate} (due to the zero plus even modes), while the running due to the brane localized matter is not affected.  This concludes our brief discussion of the beta-function coefficients in a $S^1/Z_2 \times Z_2'$ orbifold field theory.

The other crucial point for precision unification is the matching between the 5D theory and the 4D effective theory below $1/R_5$. The matching can be accounted for by the following equation where the scale $\mu$ is the matching scale~\cite{Contino:2001si}:
\begin{align}
\frac{1}{g_i^2(\mu)} = \frac{\pi R_5}{g_5^2} + \Delta_i (\mu) + \lambda_i (\mu R_5) .
\end{align}
Here $\Delta_i (\mu)$ accounts for the effect of 4D gauge kinetic operators on the orbifold fixed points, and the matching
function $\lambda_i (\mu R)$ encodes the radiative corrections from the massive KK modes once they are integrated out. Since the relative running is purely logarithmic, dimensional regularization is sufficient to get the form of the functions
$\Delta_i (\mu)$ and $\lambda_i (\mu R)$. Up to non-calculable contributions (to which we later return), we have~\cite{Contino:2001si}
\begin{align}
\lambda_i (\mu R) &= \frac{b_{\rm even}}{8 \pi^2}\left(\mathcal{I} -1 - \log[\pi] - \log [\mu R] \right) + \frac{b_{\rm odd}}{8 \pi^2} \left(- \log[2]\right) \nonumber \\
\Delta_i (\mu) &= \Delta_i(\Lambda) -\frac{b_{\rm even} + b_{\rm zero}}{8 \pi^2} \log [\mu /\Lambda]
\end{align}
where $b_{\rm even}$ and $b_{\rm odd}$ are the beta functions of the even and odd modes respectively, while
the constant $\mathcal{I} \simeq 0.02$ has been calculated numerically.
 
With the above considerations, we can obtain the relative logarithmic running of gauge couplings as a function of
the energy scale $\mu$ using dimensional regularization~\cite{Contino:2001si}
\begin{align}
\frac{4 \pi}{g^2(\mu)} &= \frac{1}{\alpha_{M_Z}} - 
   \frac{b_{\rm SM}}{2 \pi}
     \log [\frac{\mu}{M_Z}] - \frac{b_{\rm Squark}}{
    2 \pi} \log [\frac{\mu}{M_{\rm Squark}}]\nonumber\\
    &-  \frac{b_{\rm HeavyHiggs}}{
    2 \pi} \log [\frac{\mu}{M_{\rm HeavyHiggs}}] -
   \frac{b_{\rm \lambda^X}}{2 \pi}
     \log [\frac{\mu}{M_{\lambda^X}}] - \frac{b_{\rm {S}}}{
    2 \pi} \log [\frac{\mu}{M_{{S}}}]\nonumber\\
     &-\frac{b_{\rm even}}{
       2\pi} (\log[\mu/M_R] + 1 + \log [\pi] - 
         0.02) - \frac{b_{\rm odd}}{2 \pi} \log [2] +\cdots        
\label{eq:finaldiffrunning}         
\end{align}
Here $b_{\rm SM}$ is the beta function of the SM fields, $b_{\rm Squark}$ and $b_{\rm HeavyHiggs}$ are the contributions of the superpartners of the brane localized matter and the heavy higgses, respectively, $b_{\lambda^X}$ and $b_{{S}}$ are the contributions of the exotic fermion and scalar zero modes, and $b_{\rm even}$ and $b_{\rm odd}$ are the beta functions of the even and odd KK modes.  We assume all three generation squarks have an almost degenerate mass spectrum $m_{\tilde{t}} \simeq m_{\tilde{q}_{1,2}}$ in the paper for phenomenological reasons (See Section~\ref{Sec:Squarks} for more detailed discussions). In the above equation, we have added the running due to brane localized matter fields, which is the same as the corresponding MSSM fields in 4D.  For our benchmark model, the final beta functions coefficients are summarized in Table~\ref{Tab::Beta}.
 
\begin{table}
\centering
\begin{tabular}{|c|c|c|c|c|c|c|c|}
\hline 
 Group & $b_{\rm SM}$ & $b_{\rm Squark}$ &$b_{\rm HeavyHiggs}$ &$b_{\lambda^X}$ & $b_{{S}}$ & $b_{\rm even}$ &$b_{\rm odd}$  \tabularnewline
\hline 
$\SUC$ & -7 & 2 & 0 & 0 & 0 & 5 & -2  \tabularnewline
\hline 
$SU(2)_{\rm{L}}$ &-19/6 & 3/2 & 1/6 & 2/3 & 0 & 4/3 & -5/6  \tabularnewline
\hline 
$U(1)_{\rm{Y}}$ & 41/18 & 11/18 & 1/18 & 2 & 2/9 & -10/3 & 23/6 \tabularnewline
\hline 
\end{tabular}\caption{Beta function coefficients of the Standard Model fields, squarks, heavy higgses, zero mode charged fermion and scalars and even and odd KK modes.}\label{Tab::Beta}
\end{table}

\subsection{Unitarity constraint on the cutoff}\label{Sec::unitarity}

We now briefly turn to the issue of the UV scale at which a perturbative treatment must inevitably break down. 
As is well known, the scattering amplitude of gauge bosons diverges at high energies in extra dimensional theories, which implies that the theory is an effective theory which must be cut-off at a scale $M_*$ at or below the unitarity bound $\Lambda$. The tree level unitarity bound can be found by using either NDA estimates~\cite{Chacko:1999hg} or by explicitly doing the partial wave calculation of gauge boson scattering at tree level~\cite{SekharChivukula:2001hz,Chivukula:2003kq,Muck:2004br}. The scattering amplitude can be calculated as a function of the maximum number of KK modes both in momentum space~\cite{Muck:2004br} and mixed position/momentum space~\cite{Puchwein:2003jq}. Performing such a calculation
we find that the cut-off of the theory, $\Lambda$, equal, in units of $1/R_5$, to {\it twice} the maximum number of KK modes (`twice' as this sets the effective $\sqrt{s}$ at which the perturbative calculation of partial wave unitarity fails) allowed by partial wave unitarity is
\begin{align}
\Lambda \simeq \frac{64 \pi}{5 N_c g_4^2 (\Lambda)} \frac{1}{R_5} 
\end{align}
in the limit that $\Lambda R_5 \gg 1$.  Note, however, that the gauge coupling at the cut-off scale, which the dominant contribution to the scattering amplitude depends on, is quite different from the gauge coupling measured at the weak scale: The linear threshold effect between
$1/R_5$ and $\Lambda$ makes $g_4^2 (\Lambda)$ smaller than the size at $1/R_5$, and this is especially the case for the $\SUC$ gauge coupling.
We include this effect when numerically evaluating the cutoff of our theory.

\subsection{$\sin^2 \theta_W$ and the precision of $\EW$ unification}\label{Sec::su3pre}

As normal for orbifold GUTs the precision of the low-energy prediction for $\sin^2 \theta_W$ is limited by the lack
of knowledge of the coefficients of the brane-localized gauge kinetic terms that do not respect the bulk unified
symmetry~\cite{Hebecker:2001wq,Hebecker:2001jb,Hall:2001xb}.   This intrinsic uncertainty is typically of the same size as the unknown UV threshold
corrections in traditional two-loop analyses of MSSM unification, and so is not particularly problematic for the success of unification.

Utilizing the standard strong-coupling naive dimensional analysis (NDA)~\cite{Chacko:1999hg,Hall:2001xb} this intrinsic uncertainty
in the unification prediction due to the brane-localized kinetic terms can be estimated.  Assuming that the theory is strongly coupled at the fundamental scale $M_* \leq \Lambda$ for both the unified bulk gauge kinetic term and the independent brane localized gauge kinetic terms, we can obtain the size of the corresponding coupling at the matching scale $1/R_5$, which combines to produce the size of the gauge coupling in the 4D effective theory.
\begin{align}
\frac{1}{g_{ i }^2(1/R_5)} = \frac{\pi R_5}{g_5^2(1/R_5)} + \frac{1}{g_4^2(1/R_5)}
\end{align}
The brane localized gauge coupling $g_4^2(1/R_5)$ runs only logarithmically.  With the strong coupling scale $\Lambda$ not far above the compactification scale $1/R_5$, the log running does not alter the size of the overall gauge coupling as much as the linear threshold effect, especially, we expect $g_4^2(1/R_5)\simeq g_4^2 (\Lambda) \sim 16 \pi^2/N_c$. The bulk gauge coupling, $g_5$, however, runs linearly down from the scale where the theory is unified, close to the unitarity bound, to the matching scale $1/R_5$ where the combined 4D gauge coupling $g_{ i }^2(1/R_5)$ needs to be close to its weak scale value.  A sizeable bulk will ensure that such a sizeable difference can be generated~\cite{Hebecker:2001wq,Hebecker:2001jb,Hall:2001xb}.

Numerically evaluating the relative logarithmic running and threshold effects following from Eq.(\ref{eq:finaldiffrunning}) we find that the central
observed value of $\sin^2 \theta_{W}^{\overline{\rm{MS}}}(M_Z)\simeq 0.231$ is achieved when $1/R_5 \simeq 4.4 \TeV$ for our benchmark 
$\EW$ model and matter content as discussed in Section~2.1 (with corresponding $\SMLY$ beta function coefficients as displayed in Table~1).
Importantly, this value of $1/R_5$ melds nicely with that required from low-fine tuning of the EW scale in MNSUSY models~\cite{Dimopoulos:2014aua,Garcia:2015sfa}: If the value of $1/R_5$ preferred by $\SMLY$ unification had been lower than $\simeq 4 \TeV$ or much higher than this scale then either the model would have been ruled out by LHC constraints on sparticle masses, or it would have been highly tuned.  
The uncertainty in the value of $\sin^2 \theta_{W}^{\overline{\rm{MS}}}(M_Z)$ due to the brane localized kinetic terms is the same as long as unification happens before the theory hits the UV strong coupling scale.\footnote{Unification can only be discussed if the unification scale is lower than the strong coupling scale of the theory derived from the unitarity constraint of Section~2.3, which we numerically find to be $\sim 150\TeV$, in which case the precision of unification is set by $\sim N_i g_{4 i}^2/(16 \pi^2)$ regardless of the size of the bulk.} For our case, taking our central preferred value of $1/R_5 \simeq 4.4 \TeV$ and numerically evaluating the coefficients, we find the uncertainty on the $\sin^2 \theta_{W}$ prediction from 4D localized gauge kinetic terms is always less than $1\%$. 

Thus we find that precision unification of the $\SMLY$ interactions is relatively straightforward to accommodate in the MNSUSY framework
with SSSB.  We emphasize, of course, that this is not yet a full unification of SM gauge couplings, so
we now turn to the issue of unifying $\EW$ with $\SUC$.


\section{Possible $SU(6)$ unification of $\EW \times \SUC \times U(1)'$ in 6D}\label{Sec::SU6}

In this Section, we discuss the possible unification of $\EW$ with $\SUC$ into a 6D $N=4$ supersymmetric $SU(6)$ gauge theory broken by orbifold bcs to a 5D $N=2$ supersymmetric $\EW \times \SUC \times U(1)'$ gauge theory (related models were earlier
studied in Refs.\cite{Jiang:2002at,Hall:2001zb}).  We find that a {\it calculable} linear-power-law threshold effect plays a vital role.   

\subsection{The $SU(6)$ model}

We first discuss the model being considered.  We start with a pure 6D $N=4$ supersymmetric $SU(6)$ gauge theory compactified down to 5D
on an $S^1/(Z_2\times Z'_2)$ orbifold of size $\pi R_6$.  In principle the bulk theory is tightly constrained by cancellation of 6D gauge and gravitational
box anomalies but standard results~\cite{Avramis:2006nb,Green:1984bx,Schwarz:1995zw,Avramis:2005hc} show that the $N=4$ bulk theory is free of pure gauge and mixed gauge-gravity anomalies and only additional gauge singlet fields are needed to cancel the pure gravity box anomaly.   Since these
fields are gauge neutral, they do not affect gauge coupling unification.  This theory is then compactified to 5D on a $S^1/(Z_2\times Z'_2)$ orbifold in the
6th spatial direction $z$ (see the left panel of Figure~1).\footnote{It is important that this 6D to 5D orbifolding is not confused with the
independent $S^1/(Z_2\times Z'_2)$ orbifolding in the orthogonal 5th spatial direction, $y$, which has already been discussed in Section~2.}

In terms of 5D adjoint representation vector- and hyper-multiplets the bcs under the compactification are chosen to be
\begin{align}
V (x^{\mu},y, z) &= Z V (x^{\mu}, y, -z) Z^{-1} = Z' V (x^{\mu}, y, -z+ 2 \pi R_6) Z'^{-1} \nonumber\\
\Phi (x^{\mu},y, z) &=   Z \Phi (x^{\mu}, y, -z) Z^{-1} = - Z' \Phi (x^{\mu}, y, -z + 2 \pi R_6) Z'^{-1} 
\end{align}
where ${Z,Z'} = {\diag(1,1,1,-1,-1,-1),\diag(1,1,1,1,1,1)}$.  In other words the orbifold bcs for the bulk gauge multiplet are
\begin{align}
& V : \begin{pmatrix}
 (+,+)_{3 \times 3} & (-,+)_{3 \times 3}  \\
(-,+)_{3 \times 3} & (+,+)_{3 \times 3}
\end{pmatrix}
& \Phi : 
\begin{pmatrix}
(-,-)_{3 \times 3}  & (+,-)_{3 \times 3} \\
(+,-)_{3 \times 3} & (-,-)_{3 \times 3} .
\end{pmatrix}
\end{align}

These bcs preserve $N=2$ SUSY and $\EW \times \SUC \times U(1)'$ gauge symmetry in the low energy effective theory (correspondingly the zero mode spectrum is a $N=2$ vector multiplet of $\EW \times \SUC \times U(1)'$).
In more detail, the $z=\pi R_6 $ fixed point preserves the $SU(6)$ gauge symmetry while breaking the $N=4$ supersymmetry to $N=2$, while the $z=0$ fixed point is a $N=2$ supersymmetric $\EW \times \SUC \times U(1)'$ gauge theory where the $\EW$-extended higgs and lepton hypermultiplets of Section~2 live -- see the left panel of Figure~1.    (The $U(1)'$ gauge symmetry is later broken at scale $1/R_5$. In Section~\ref{Sec::U1prime}, we will discuss the case where we identify this $U(1)'$ as the extended gauge structure needed to raise the higgs pole mass to its observed value.)

It will be important for our analysis that the {\it bulk} field content after the compactification has an accidental $Z_2$ symmetry that interchanges the two $SU(3)$ gauge groups, which is only broken by matter fields localized on the $z= 0$ brane. The feature that the three flavors of lepton fields and the higgs fields are the only source of breaking of the accidental $Z_2$ symmetry is a crucial ingredient that leads to precision predictions for unification of $\EW \times \SUC$.

We now turn to a discussion of the calculable differential power-law {\it threshold} effect that splits the 
$\EW \times \SUC$ gauge couplings at low energy starting from the $SU(6)$-dominated unified value at high energies.

\subsection{Calculable differential linear threshold effects}

The counter-term for the bulk 6D gauge coupling has dimension $\rm{[mass]}^2$ and therefore can in principle have linear and quadratic UV cut-off $\Lambda$ dependence.  However, in the present model, the bulk is exactly $SU(6)$ symmetric and $N=4$ supersymmetric, and so there is no differential quadratic cut-off dependent threshold effect.  In addition, in our model the KK mode spectrum of the 6D bulk fields respect an accidental $Z_2$ symmetry that interchanges the two $SU(3)$ gauge groups, and so there are no differential effects between the two $SU(3)$ gauge
couplings due to 6D bulk supermultiplets at all. 

In fact, the leading differential correction in our model is a linear threshold entirely due to the 5D fields localized on the
$z= 0$ brane which do not respect the $Z_2$ symmetry between the two $SU(3)$ gauge groups. Remarkably these linear threshold
effects can be calculated in a $N=2$ supersymmetric gauge theory with the results of Intriligator, Morrison, and Seiberg~\cite{Seiberg:1996bd,Intriligator:1997pq} as we now briefly explain (we closely follow the discussion of~\cite{Hebecker:2004xx}).

Power-like threshold corrections to gauge couplings are a generic feature of higher dimensional gauge theory. 
Fortunately for us, these corrections are very constrained by the combined gauge symmetry and supersymmetry
of the 5D theory under consideration.
In the 4D superfield language of Section~2.1 the low energy description of the 5D SUSY theory, specifically the two-derivative terms that determine the low-energy gauge couplings, is fully characterized by a holomorphic pre-potential $\mathcal{F}(\Sigma)$ where $\Sigma=\Sigma^a T^a$ is the adjoint representation 4D chiral supermultiplet contained in the 5D vector supermultiplet:
\begin{align}
\mathcal{L} = \frac{1}{2}\int {\rm d}^4\theta \frac{\partial \mathcal{F}(\Sigma)}{\partial \Sigma^a} ({\bar \Sigma} e^{2V})^a +  \frac{1}{2}\int {\rm d}^2\theta \frac{\partial^2 \mathcal{F}(\Sigma)}{\partial \Sigma^a \partial \Sigma^b} W^a W^b  + h.c.
\end{align}
(here the dependence of all the fields on the 5th-dimension coordinate is simply treated as an extra parameter).
Vitally, the pre-potential is consistent with 5D gauge invariance only if it is at most cubic~\cite{Seiberg:1996bd,Intriligator:1997pq}\footnote{The $n>3$ terms in the pre-potential will lead to, on the Coulomb branch, bosonic Lagrangian terms in the form of $\Sigma^{n-3} A_5 {\rm d} A \wedge {\rm d} A$ in terms of 4D gauge fields and scalars, or with 5D gauge fields $\Sigma^{n-3} A \wedge {\rm d} A \wedge {\rm d} A$.
The $A \wedge {\rm d} A \wedge {\rm d} A$ piece is the 5D Chern-Simons term, which under gauge transformation become a total derivative. Therefore, only when $n \leq 3$ is $\Sigma^{n-3} A \wedge {\rm d} A \wedge {\rm d} A$ gauge invariant. Similar arguments apply to 6D $N=2$ theories, with the result that $n\leq 4$, the corresponding Lorentz invariant Lagrangian term being $A_5 A_6 {\rm d} A \wedge {\rm d} A$.
This term, upon compactification, might lead to corrections to the low energy effective theory at two derivative level. If both $A_5$ and $A_6$ are taken to be along the direction of $\SUC \times \EW \times U(1)'$, after compactification, the low energy 5D $\SUC \times \EW \times U(1)'$ gauge symmetry and $N=2$ supersymmetry implies this term with dimension six is absent in the low energy theory. If $A_5$ and $A_6$ are both taken to be along the direction of $SU(6) / \SUC \times \EW \times U(1)'$, there will in principle be corrections at two derivative level to the gauge coupling. However, due to the accidental $Z_2$ symmetry of the bulk spectrum, these contributions are universal for the two $SU(3)$ and does not introduce non-calculable differential threshold effect.} and this together with holomorphy and the restrictions of 5D Lorentz invariance strongly constrains the exact form
of the quantum pre-potential.  This feature is important because the unification with calculable power-like threshold corrections might be spoiled by non-calculable contributions from higher dimensional operators suppressed by $1/M_*$. In the case where the fundamental scale $M_*$ is comparable to the unification scale, these operators might totally destroy the unification prediction.  However the form of the pre-potential, and thus all two-derivative terms is determined by the results of Refs.~\cite{Seiberg:1996bd,Intriligator:1997pq,Hebecker:2004xx}, and so the higher-dimension operators that might otherwise be dangerous for predictable unification are eliminated.

For example, the pre-potential at the classical level is
\begin{align}
\mathcal{F}^{(cl)}(\Sigma) = \frac{1}{4 g^2_{5,cl}} \delta_{ab} \Sigma^a\Sigma^b + \frac{c_{cl}}{48 \pi^2} d_{abc} \Sigma^a \Sigma^b \Sigma^c ,
\end{align} 
with $d_{abc} \equiv \frac{1}{2} {\rm Tr} (T_a \{T_b,T_c\})$, leading to the classical two derivative bosonic Lagrangian terms
\begin{align}
\mathcal{L} \supset \frac{1}{4 g_{5,cl}^2} {\rm Tr} F^2 + \frac{c_{cl}}{16 \pi^2} {\rm Tr} \Sigma F^2 ,
\end{align} 
where the second term is the 'classical' Chern-Simons (CS) term. After including quantum corrections, the pre-potential is constrained to have the {\it exact} form~\cite{Seiberg:1996bd,Intriligator:1997pq,Hebecker:2004xx}
\begin{align}
\mathcal{F}(\Sigma) =\mathcal{F}^{(cl)}(\Sigma)+ \frac{1}{96 \pi^2} \left(\sum_{\alpha} |\alpha_a \Sigma^a|^3 - \sum_f \sum_{\lambda} |\lambda_a \Sigma^a + m_f|^3 \right).
\label{eq:exactprepot}
\end{align} 
Here the sum runs over the roots of the unbroken gauge group $\alpha$ and the weights of the relevant matter representation $\lambda$, and the modulus signs only determine whether the cubic terms are to be multiplied by $\pm 1$, so $\mathcal{F}(\Sigma)$ remains locally holomorphic. The last two terms come from integrating out the gauge bosons in the broken directions and the heavy 5D matter fields with mass $m_f$. The corresponding effective 5D gauge coupling in the quantum theory is

\begin{align}
\frac{1}{4 g_{5,eff}(\tilde{\Sigma})^2} =\frac{1}{4 g^2_{5,cl}} + \frac{c_{cl}}{16 \pi^2}  \tilde{\Sigma}^a + \frac{1}{16 \pi^2} \left(\sum_{\alpha} \alpha_b^2|\alpha_a \tilde{\Sigma}^a| - \sum_f \sum_{\lambda} \lambda_b^2 |\lambda_a \tilde{\Sigma}^a + m_f| \right),
\label{eq:exactcoupling}
\end{align} 
where $\tilde{\Sigma}$ is the vev of the scalar component of the chiral adjoint field. It is important for us that this formula gives the exact dependence of the gauge coupling on the masses of the hypermultiplets $m_f$.
Note that integrating out $n_f$ 5D hypermultiplets with masses $m_f \rightarrow \pm \infty$ in the fundamental representation induces a shift in the effective CS term $c_{cl} \rightarrow c_{cl} \mp 
\frac{n_f}{2}$. This can be seen because for $m_f \gg |\Sigma^a|$, the sign of $m_f$ defines the sign of the modulus of the final term, and expanding out the final term in linear order in $\tilde{\Sigma}$ effectively shifts $c_{eff} = c_{cl} \mp \frac{n_f}{2}$. 

The additional term linear in the mass of the hypermultiplet $m_f$ in the limit $m_f \rightarrow \pm \infty$ does not change the relative size of the gauge coupling between the strong coupling scale and the scale $1/R_5$. This is because at the strong coupling scale $M_*$ (which we later take to be the unification scale $M_{\rm{GUT}}$), it is the effective coupling $g_{5,eff}$ given by
\begin{align}
\frac{1}{4 g_{5,eff}^2} =\frac{1}{4 g^2_{5,cl}} - \frac{1}{32 \pi^2}\sum_f  |m_f|
\label{eq:strongcoupling}
\end{align} 
that goes to the strong coupling value. 
Thus the unitarity constraint on the model is not affected as heavy fermions are added to the theory, and can be calculated with knowledge of the low energy spectrum of the theory and the corresponding 'classical' CS-term coefficient of the low energy effective theory.

In our setup, the $SU(6)$ symmetry breaks down by boundary conditions to $\SUC \times \EW \times U(1)'$ with an accidental $Z_2$ symmetry interchanging $\SUC$ and $\EW$. We define our UV theory by demanding that the only breaking of the $Z_2$ symmetry in the 5D effective theory is a soft breaking due to the N=2 SUSY preserving masses of the brane localized hypermultiplets on the $z=0$ brane.\footnote{For example, we can preserve the $Z_2$ symmetry in the UV by having 5 pairs of triplet hypermultiplets under both $\EW$ and $\SUC$, but then softly break the $Z_2$ by taking masses 0 and $M_0$  for the 5 pairs of $\EW$ and $\pm M_0$ for the 5 pairs of $\SUC$. Then below $M_0$, the heavy hypermultiplets are integrated out and the low lying $\EW$ hypermultiplets $(\mathcal{L},\mathcal{H}_{u,d})$ are the only $Z_2$ breaking matter content. Additional $Z_2$ symmetric matter content above $M_0$ will not change the differential threshold effect.}

The theory below the $Z_2$ breaking scale $ M_0$ has a 'classical' CS coefficient of $\delta c_{cl} = 0$ and $+ \frac{5}{2}$ for $\SUC$ and $\EW$, respectively. In the following, we will account for the $Z_2$ breaking linear threshold corrections in the low energy effective theory with the classical CS terms.

Since the only $Z_2$ breaking effect is the soft breaking due to the masses of the hypermultiplets (ignoring the gauge kinetic term on the $y = \pi R_5$ brane), the differential linear threshold effect between the 5D $\EW$ and $\SUC$ gauge couplings at the matching scale $1/R_6$ is 

\begin{align}
\frac{1}{ g^2_{5,\rm{EW}}\left(1/R_6 \right)}  - \frac{1}{ g^2_{5,\rm{C}}\left(1/R_6 \right)} =  \frac{5/2}{4 \pi^2} M_{0} .
\end{align} 

Here we use the fact that the $Z_2$ symmetry ensures that the gauge coupling and CS coefficients above the $Z_2$ breaking scale $M_0$ are the same. 

It should be noted that there will also be logarithmic differential running between the two $SU(3)$'s, however, the logarithmic differential running and finite threshold corrections can be understood with 4d KK modes summation and therefore is already taken into account with methods introduced in Section~\ref{Sec::difflog}. 
  
\subsection{The precision of $SU(6)$ unification and $\alpha_3(M_Z)$}

The $Z_2$ symmetry on the $z=0$ brane ensures that the $SU(6)$ breaking 5D gauge kinetic term is the same for $\EW$ and $\SUC$. However, there are $Z_2$ breaking 5D gauge kinetic terms on the $y=\pi R_5$ brane. The 5D gauge kinetic terms, like the 4D gauge kinetic term discussed in Section \ref{Sec::su3pre} will generate an non-calculable $Z_2$ breaking contribution to the gauge coupling at low energies, and therefore affect the precision of unification.

The unitarity constraint on the size of the sixth dimension and the precision of unification can be estimated with NDA~\cite{Chacko:1999hg}. Taking into account all powerlaw and logarithmic 1-loop effect, we find, with a fundamental scale at $ M_* \sim 120\TeV$, the bound on the size of the sixth dimension is $M_* R_6 \leq 3$. Then, the size of the non-calculable threshold correction due to the (volume diluted) $Z_2$ breaking kinetic terms on the $y = \pi R_5$ brane is approximately $5\%$ as follows from a standard strong coupling analysis. It should be noted the size of the non-calculable contributions does not directly translate to the precision of the prediction of gauge couplings at the electroweak scale. Combined with the estimate of the size of the operators localized on the 4D fixed point, the precision on the $\sin^2 \theta_{\rm W}$ prediction is $2 \%$ (fixing $\alpha_3$ and $\alpha_{\rm EM}$), while the precision of the prediction of $\alpha_3$ is $7 \%$ (fixing $\sin^2 \theta_{\rm W}$ and $\alpha_{\rm EM}$), signaling the unusual two step unification in this model.


\section{$U(1)'$ and the higgs mass}\label{Sec::U1prime}

\subsection{$U(1)'$ symmetry breaking and anomalies}
The $U(1)'$ from the broken $SU(6)$ symmetry can be used to lift the higgs pole mass to $126 \GeV$ by choosing proper $U(1)'$ charge of the higgs and the fields $\phi_{\pm}$, and 5D bcs of the $U(1)'$ gauge boson. The higgs $H_{u,d}$ must have opposite $U(1)'$ charge so that the brane localized divergent Fayet-Iliopoulos term vanishes~\cite{Barbieri:2002ic,Marti:2002ar,Barbieri:2001cz}. The bcs for $\phi_{\pm}$ that lead to a non-decoupling D-term~\cite{Maloney:2004rc} is
\begin{align}
 \phi_{\pm} &:& \begin{pmatrix}
 (+,-) \cr (+,-) \cr (+,-)
 \end{pmatrix}, ~~
\phi_{\pm}^c &:& \begin{pmatrix}
 (-,+) \cr (-,+) \cr (-,+)
 \end{pmatrix}, ~~
\tilde{\phi}_{\pm} &:& \begin{pmatrix}
 (+,+) \cr (+,+) \cr (+,+)
 \end{pmatrix}, ~~
\tilde{\phi}_{\pm}^c &:& \begin{pmatrix}
 (-,-) \cr (-,-) \cr (-,-)
 \end{pmatrix}.
\end{align}\label{Eq:phiboundary}

The brane localized superpotential
\begin{align}
W \supset \int {\rm d} y {\rm d} z \frac{\lambda}{M_*} X (\phi_+ \phi_- - v_{\phi}^3)\delta(y - \pi R_5)\delta (z),
\end{align}
with bcs in Eq.(\ref{Eq:phiboundary}), generates both a vev for $\phi_{\pm}$ to break the $U(1)'$ gauge symmetry and an F-term for $X$ around the size of the fifth dimension $1/R_5$ to get Standard Model Yukawa couplings~\cite{Dimopoulos:2014aua}. The gauge coupling $g'$ is normalized as $g'^2 = \frac{1}{3} g_c^2$ at the GUT scale. The differential linear threshold effect between the $U(1)'$ and $\SUC$ dominates the differential logarithmic running, and the $U(1)'$ linear threshold beta function is $b_{U(1)'} = (33/4 + 2 N_{\phi}^2)/3$, where $N_{\phi}$ is the $U(1)'$ charge of pairs of $\phi_{\pm}$ fields. Starting from a strongly coupled theory in the far UV, the low energy gauge coupling is approximately
\begin{align}
g'^2 \simeq \frac{1}{3} \frac{4 \pi^2}{b_{U(1)'} (M_* R_5)} \simeq 0.1
\end{align}
for $N_{\phi}=1$. Here, the $Z_2$ symmetry does not ensure a vanishing relative brane localized gauge kinetic term on the $z=0$ brane, which generates the main uncertainty on the prediction of $g'$. The uncertainty in the size of $g'$ and the $\phi_{\pm}$ field vev $v_{\phi}$ contribute a major uncertainty in the prediction of the $U(1)'$ gauge boson mass and the mass of the Higgs. With a more careful treatment of the differential linear threshold effect and logarithmic running, the resulting $U(1)'$ gauge boson mass is
\begin{align}
M_{U(1)'}^2  \simeq 2 g' N_{\phi} v_{\phi_{\pm}}^{3/2} \left(\frac{1}{\pi R_5}\right)^{1/2} \approx (2.4 \pm 0.2 \TeV )^2 \left(\frac{N_{\phi}}{1}\right)\left(\frac{v_{\phi}}{4.4 \TeV}\right)\left(\frac{g'}{0.26}\right)
\end{align}

The broken $U(1)'$ symmetry is not anomalous in 5 dimensions but has 4D anomalies with the $U(1)'$ charge assignment of the lepton $\mathcal{L}$ and the right-handed quark $u^c, d^c$. The 4D anomalies of this broken symmetry suggest that localized fermions needed to be added on the $y= \pi R_5, z=0$ or $y=0,z=0$ brane to UV complete the theory~\cite{Preskill:1990fr}. The mass of these states need to be lighter than
\begin{align}
m \approx 4 \pi \frac{M_{U(1)'}}{g'} \simeq 150 \TeV
\end{align}
It is not inconceivable that the mass of the localized states is close to the unification scale and does not significantly change the prediction of the $U(1)'$ gauge coupling at the electroweak scale. Indeed, the uncertainties on the predictions of the physical higgs mass due to the uncertainty of the masses of the heavy states are minor compared with other uncertainties.

\subsection{Higgs mass from non-decoupling D-term}

\begin{figure}[htb]
\centering
\includegraphics[width=0.8\textwidth]{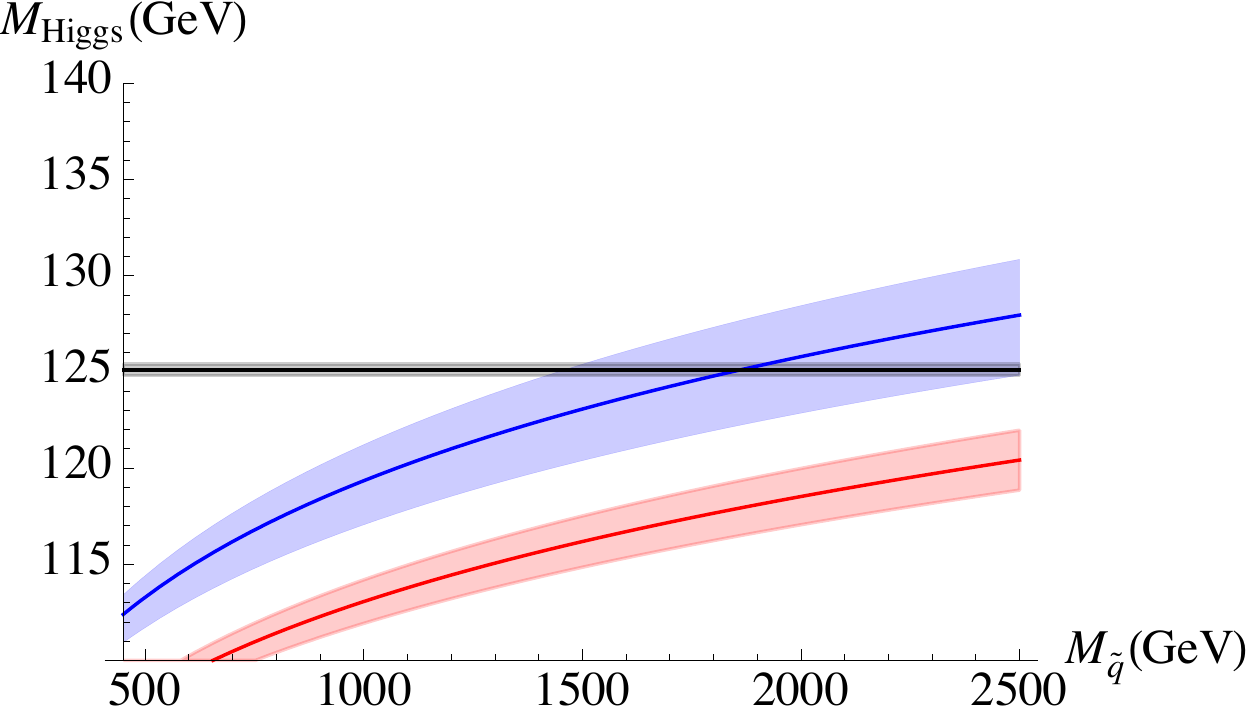}
\caption{The physical higgs mass as a function of the squark (stop) masses with $1/R_5 = 4.4 \TeV$. The blue curve is the prediction for the physical higgs mass with the contributions from the non-decoupling D-term of $U(1)'$ and the stop loop, with the light blue band the theoretical uncertainties on the prediction. The red curve is the prediction for physical higgs mass with the contributions from the the stop loop alone, with the light red band the theoretical uncertainties on the prediction. The black line and the corresponding band is the measured value~\cite{Aad:2015zhl}.}\label{Fig::higgsmass}
\end{figure}

The contribution to the higgs quartic from the non-decoupling D-term can be written as~\cite{Garcia:2015sfa}
\begin{align}
m_{h}^2 \simeq 2 \left(\frac{1}{4} \left(g_1^2 +g_2^2\right) +  2\delta \lambda_{U(1)'} +\delta \lambda_{\rm stop \, LL} + \delta \lambda_{\rm EW\, 1-loop} + \cdots \right) v_{\rm EW}^2,
\end{align}
where $\delta \lambda_{U(1)'} = \frac{g'^2}{2} N_{\phi}^2 f(M_{U(1)'} R)$ and $v_{\rm EW} = 174 \GeV$.  Here $f(M_{U(1)'} R)$ is a function of the $U(1)'$ gauge boson mass and the size of the fifth dimension~\cite{Dimopoulos:2014aua,Garcia:2015sfa} that determines the relative suppression on the $U(1)'$ contributions to the masses of the higgs boson due to decoupling.  We include the all-orders leading-log contributions from the stop and fixed-order 1-loop leading-log contributions from the EW states to the quartic coupling of the Higgs~\footnote{See~\cite{Garcia:2015sfa} for more details.}. The physical Higgs mass with the benchmark values (see table \ref{Tab::mass}) for the mass parameters is
\begin{align}
m_{h} \simeq  123.7^{+2.7}_{-2.5} \GeV,
\end{align}
with $U(1)'$ charge $N_{\phi} = 1$, and the mass of the top quark $m_{t} = 173.2 \pm 0.9 \GeV$. The comparison between the prediction of the physical higgs mass with the compactification scale $1/R_5 = 4.4 \TeV$ and the mass measured by ATLAS and CMS collaboration~\cite{Aad:2015zhl} is shown in Figure~\ref{Fig::higgsmass}. The uncertainty on the higgs mass mainly comes from the theoretical uncertainty in the loop contributions from the stop and the experimental uncertainties on the top quark mass measurement ($ \sim 2\GeV$), and the uncertainties on the prediction of $U(1)'$ coupling $g'$ and $\phi_{\pm}$ vev $v_{\phi}$ ($ \lesssim 1.5\GeV$). There are additional contributions to the higgs quartic coupling from electroweak box diagram with $S_u$ and $A^X$, and $\tilde{S}_u$ and $\lambda^X$, shifting the physical higgs mass by a negligible $0.05\GeV$.
   
\section{Phenomenology of unified MNSUSY}

The requirement that the theory unifies into a $SU(6)$ in six dimension favors certain relationships between the brane localized squark and heavy higgs masses, the fifth dimension size, $1/R_5$, and the zero mode gauginos $\lambda^X$ and singlet higgs ${S}_{u,d}$ masses. In this Section, we discuss the current experimental probes of the light modes in the theory and their implications to the spectrum of the unified theory. We will focus on the case where the unified 6D theory is UV completed into a theory with a large gravitational bulk of dimension higher than 6, and assume that the light states in the theory will eventually decay into bulk LSPs~\cite{Dimopoulos:2014psa}.

\subsection{Auto-concealment and limits on the squark masses}\label{Sec:Squarks}

The brane localized squarks get masses from higher dimensional operators in the ${\rm K\ddot{a}hler}$ potential in Eq.(\ref{Eq:Mscalars}). Since all three generation of squarks are on the brane, it is most natural to have their masses near degenerate. In this case, the first two generation squarks decay to jets plus bulk states while the stop decays to top quark plus bulk states. These decays are studied in~\cite{Dimopoulos:2014psa} for the case where the stop decay is prompt. The decay width of the squarks is estimated to be~\cite{Dimopoulos:2014psa}
\begin{align}
\Gamma \simeq  \frac{\Omega_d m_{\tilde{q}}^{d+3}}{ 8 \pi (2\pi)^d M_*^{d+2}} \int_0^1 x^{d+1} (1-x^2)^2 {\rm d} x,
\end{align}
where $m_{\tilde{q}}$ is the squark mass. With the benchmark value $m_{\tilde{t}} = 800 \GeV$ and $M_* = 120 \TeV$, the squark and stop decay promptly when the number of large gravitational dimension $d=3$, decay with a displaced vertex when $d=4$ and are long-lived when $d \geq 5$. The experimental limit on the squark with three large gravitational dimension is around $450 \GeV$~\cite{Dimopoulos:2014psa}. When the first two generation sqaurks have significant different mass compared with the stop, to be more specific, $|\delta m| = |m_{\tilde{t}} - m_{\tilde{q}_{1,2}}| > m_{t}$, the constraint is more stringent due to signatures from multiple top quarks from the cascade decay. Throughout the paper, we assume $m_{\tilde{t}} \simeq m_{\tilde{q}_{1,2}}$.

\subsection{Mass limits on the zero mode gauginos and singlet higgs}

The masses of the singlet higgs and gauginos come from the superpotential and ${\rm K\ddot{a}hler}$ potential terms in Eqs.(\ref{Eq:Mscalars}) and (\ref{Eq:Mgaugino}) respectively.  The mass of the gauginos ${\lambda^X}$ is $\sim 2 \TeV$ when the size of the fifth dimension $1/R_5 \sim 4 \TeV$, while the mass of the singlet higgs ${S}^c_{u,d}$ is less than $\sim 500 \GeV$.

\begin{figure}[htb]
\centering
\begin{minipage}[t]{0.4\textwidth}
\includegraphics[width=\textwidth]{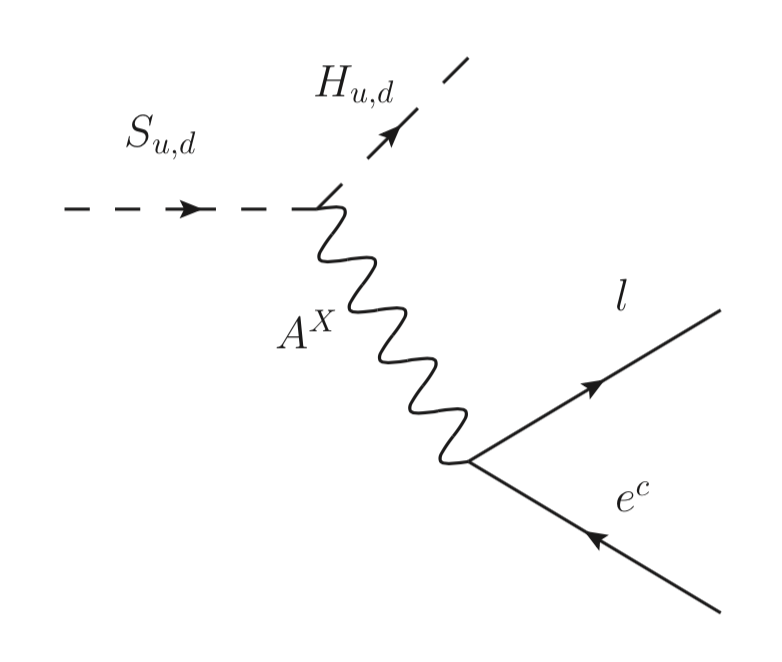}
\end{minipage}
\begin{minipage}[t]{0.5\textwidth}
\centering
\includegraphics[width=\textwidth]{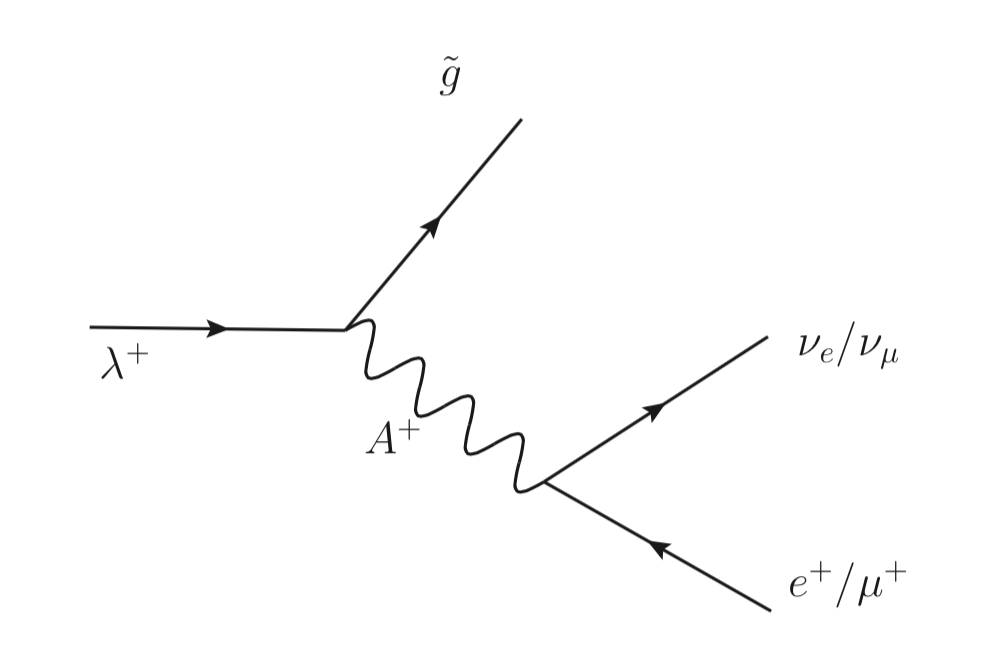}
\end{minipage}
\caption{{\it Left panel:}  The singlet higgs $S_{u,d}$ decay through heavy gauge boson to lepton pair. {\it Right panel:} The displaced decay of singly charged gaugino $\lambda^+$ to charged leptons and missing energy.}\label{Fig::scalar}
\end{figure}

These light charged particles are potentially observable in the LHC.  The singlet higgs ${S}_{u,d}$ couples to the SM fields only through $\EW$ gauge couplings. The production cross section of the singlet higgs $S_{u,d}$ is twice the production cross section of the right handed stau with all gauginos decoupled. These singlet higgs decay through off-shell gauge boson $A^X$ to $H_{u,d}$ and dilepton as shown in Figure~\ref{Fig::scalar}. A single event from $S_{u,d}$ decay has $\sim 20 \%$ chance to have four or more first two generation leptons with significant missing energy. The two leptons, $l$ and $e^c$, will be same-sign same-flavor dilepton if $A^X$ is doubly charged. The chance of having two same-sign same-flavor first two generation leptons, jets and significant missing energy in the decay of $S_{u,d}$ is also approximately $20 \%$. The stringent constraint on their masses comes from multi-lepton~\cite{Aad:2014iza,Chatrchyan:2014aea} and same-sign dilepton~\cite{Aad:2014pda,Chatrchyan:2013fea} searches, and the limit can be estimated to be $350 \GeV$. These singlet higgs, acquiring their masses through similar higher dimensional operators as the heavy higgses in MSSM, is most likely to get a mass comparable to the heavy higgses $H_d$. The coefficient $c_{H_{u,d}}$ and $c_{S_{u,d}}$, however, are not related by $\EW$ gauge symmetry and therefore it is not inconceivable that the doublet and singlet has a mass splitting large enough for the decay to be prompt, for example, contributions from the electroweak symmetry breaking. The collider signature of the case where $S_{d}$ is near degenerate with $H_d$ depends highly on how $H_d$ decays. We will not attempt to provide an in-depth study of that scenario.

\begin{figure}
\centering
\includegraphics[width=0.4\columnwidth]{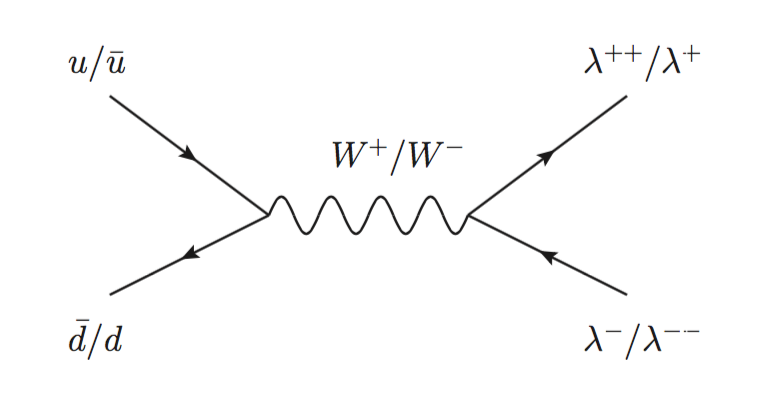}
\includegraphics[width=0.4\columnwidth]{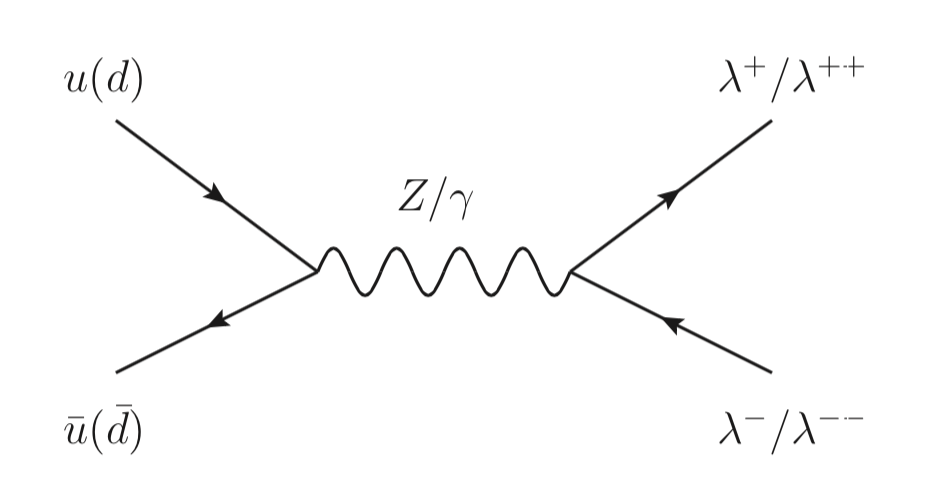}
\caption{The leading LHC production processes of the singly and doubly charged gauginos $\lambda^X$.}\label{Fig::production}
\end{figure}

The gaugino $\lambda^X$ forms an $SU(2)_{\rm L}$ doublet containing a pair of doubly charged fermion($\lambda^{\pm \pm}$) and a pair of singly charged fermions ($\lambda^{\pm}$) under $U(1)_{\rm EM}$.  These $\lambda^X$ gauginos get produced with similar cross sections as the MSSM electro-weakinos~\cite{Beenakker:1999xh,Choi:1998ut,Choi:1998ei,Kramer:2012bx} through both charged current and neutral current processes as shown in Figure~\ref{Fig::production}. 

The splitting between the two fermions, $\lambda^{\pm \pm}$ and $\lambda^{\pm}$, is equal to the usual splitting between the charged and neutral winos in anomaly mediated supersymmetry breaking (AMSB) scenarios~\cite{Randall:1998uk,Giudice:1998xp}, with one loop splitting~\cite{Ibe:2012sx,Cirelli:2005uq} between $\lambda^{\pm \pm}$ and $\lambda^{\pm}$ being $\approx 1.01\GeV$.  This splitting leads to a prompt decay of $\lambda^{\pm \pm}$ to $\lambda^{\pm}$ and a charged pion. The singly charged gauginos $\lambda^{\pm}$ are the lightest supersymmetric particle along the broken directions of $\EW$ and therefore decay to the true LSP states in the gravitational bulk. The dominant channel for them to decay is through processes shown in Figure~\ref{Fig::scalar}. The decay width of these particles is~\cite{Dimopoulos:2014psa}
\begin{align}
\Gamma \sim \frac{g_{\EW}^2}{(4 \pi)^3} \frac{m_{\lambda^X}^{5}}{{1/2R}^4} \frac{\Omega_d m_{\lambda^X}^{d+2}}{(2\pi)^d M_*^{d+2}} \int_0^1 x^{d+1} (1-x^2)^2 {\rm d} x,
\end{align}
The lifetime of $\lambda^{\pm}$ is approximately $1 - 50\,{\rm ns}$ for $d=3$, while $\lambda^{\pm}$ become relatively collider-stable for $d > 3$.  Such collider-stable $\lambda^+$ is constrained by searches for stable charged particles in LHC~\cite{ATLAS:2014fka,Khachatryan:2015lla}. Recasting the bound on chargino pair production suggest gauginos, $\lambda^X$, with masses smaller than $1\TeV$ are excluded. For $d=3$, the collider signature is a mixture of kinked vertices~\cite{Asai:2011wy,Graham:2012th} and long-lived charged particles depending on the mass of the gaugino $\lambda^X$ and the fundamental scale $1/M_*$. Searches in CMS~\cite{CMS:2014gxa,Khachatryan:2015lla,Khachatryan:2014mea} and ATLAS ~\cite{Aad:2013yna,ATLAS:2014fka} limits the mass of the charginos to be heavier than $750\GeV$ for $\lambda^{\pm}$ with lifetime longer than $1\,{\rm ns}$.

The decay of these gauginos $\lambda^X$ and singlet higgs $S_{u,d}$ produces very distinct signatures in the LHC including kinked charged tracks, high $p_T$ same-sign same-flavor di-lepton pair with significant missing energy.  Thus we believe that LHC 14 has a good coverage over most of the natural parameter space of the model that explains both EW unification and the higgs mass.  A detailed analysis of possible search optimizations is beyond the scope of this paper.

\subsection{$U(1)'$ gauge boson}

The $U(1)'$ gauge boson is produced at the LHC with cross section comparable to that of standard $Z'$ gauge bosons. Depending on the size of the fifth dimension $1/R_5$, their masses and couplings to the Standard Model quarks and leptons vary from $2.5 \TeV$ to $3.5 \TeV$ and $0.2$ to $0.3$, respectively. Since the $U(1)'$ gauge coupling is much smaller than the $\EW$ gauge coupling at $1/R_5$, the cross section for production of the $U(1)'$ gauge boson is approximately $ 40\%$ to $80\%$ (for $g'=0.2$ and $0.3$, respectively) of the production cross section of a $Z'_{SM}$ with the $SU(2)_{\rm L}$ gauge coupling.  The right-handed quarks have twice the $U(1)'$ charge compared with the leptons while the left-handed quarks are not charged under $U(1)'$.  As a result, the $U(1)'$ gauge boson has $16/27$ branching ratio to decay into first two generation quarks and $2/81$ branching ratio to decay into either electron or muon pairs. Reinterpreting the constraints from searches for di-jet resonance and di-lepton resonance gives a conservative limit on $Z'$ mass around $2.3 (2.5) \TeV$~\cite{Khachatryan:2014fba,Aad:2014cka,Khachatryan:2015sja,Aad:2014aqa} for $g'=0.2(0.3)$.  Current LHC searches for heavy resonances do not yet constrain the most interesting parts of parameter space of our model, but future searches for di-lepton resonances should provide a useful reach.

\subsection{Proton decay and neutron anti-neutron oscillation}\label{sec::protondecay}

Proton decay is a generic prediction of grand unified theories~\cite{Georgi:1974sy}, and the non-observation of proton decay puts strong constraints on the form of the unified theory, the constraint being especially severe when the quantum gravity scale is low~\cite{ArkaniHamed:1999dc}. Given that the fundamental scale is bounded to be around 100 TeV in our model due to unitarity considerations, operators with dimension 4,5,6 are all dangerous and may lead to too fast proton decay. However, the $U(1)'$ gauge symmetry can be used to suppress dangerous proton decay operators: With the charge assignment for the $U(1)'$ in Table~\ref{Table::R}, the proton decay operator is forbidden due to the broken $U(1)'$ gauge symmetry. If the field $\phi_{\pm}$ is integer charged under $U(1)'$, the dangerous proton decay operator vanishes since $L, e^c$ are the only half-integer charged fields under the $U(1)'$ symmetry.

Moreover, the neutron anti-neutron oscillation operators~\cite{Mohapatra:2009wp,Babu:2013yww} that are unique in SUSY theories, {\it i.e.}, the superpotential operators, can be eliminated by unbroken R-symmetry on the $z=0, y=\pi R_5$ brane. ${\rm K\ddot{a}hler}$ operators localized on $z=0, y=\pi R_5$ brane might lead to neutron anti-neutron oscillation if the $U(1)'$ charge of $\phi_+$ is $N_{\phi} = 1, 2$. For example, if $N_{\phi} = 1$
\begin{equation}
K \supset \int {\rm d}^4\theta \delta (y- \pi R_5) \delta (z) \frac{c_{12}}{M_*^{8}} QQQQ d^{c \dagger} d^{c \dagger} X^{\dagger} \phi_+ \phi_+
\end{equation}
Once supersymmetry is broken, this leads to the operator
\begin{equation}
\mathcal{L} \supset  \frac{c_{12} }{M_*^{8}}Q Q \tilde{Q} \tilde{Q} \tilde{d^{c \dagger}}  \tilde{d^{c \dagger}}  F_X^{\dagger} v_{\phi^+}^3 + ...
\end{equation}
Neutron anti-neutron oscillation happens when the squark loops are completed with gauginos and leads to the dimension 9 operator similar to that of the Standard Model. Such an operator produces neutron anti-neutron oscillation at the rate
\begin{equation}
\delta m \simeq 10^{-41} \GeV \left(\frac{c_{12}}{1}\right) \left( \frac{120 \TeV}{M_*}\right)^{8} \left(\frac{v_{\phi_+}}{5 \TeV}\right)^3 \left(\frac{\Lambda_{QCD}}{200\, {\rm MeV}}\right)^6
\end{equation}
and given current bound from the ILL reactor experiment~\cite{BaldoCeolin:1994jz}, and comparable bounds from nucleon decay experiments~\cite{Takita:1986zm,Chung:2002fx,Kopeliovich:2011aa,Abe:2011ky}, of $\delta m = 6 \times 10^{-33} \GeV$, the theory is safe from neutron anti-neutron oscillation constraints. 

Finally, the higher dimensional operators that lead to proton decay and neutron anti-neutron oscillation could be generated by weak quantum gravity effect alone, but they can receive extra, exponential, suppression in some string theories into which the unified theory UV completes.  We note that in our setup where the quarks are only charged under the SM group and the leptons only charged under $\SUC \times \EW$, the calculable gauge boson or gaugino mediated proton decay is totally absent.  Therefore, though additional assumptions on the UV completion might in some cases be required, it is still conceivable that the minimal model shown in Figure~\ref{Fig::su6} does not lead to proton decay.

\begin{figure}
\centering
\includegraphics[width=0.32\columnwidth]{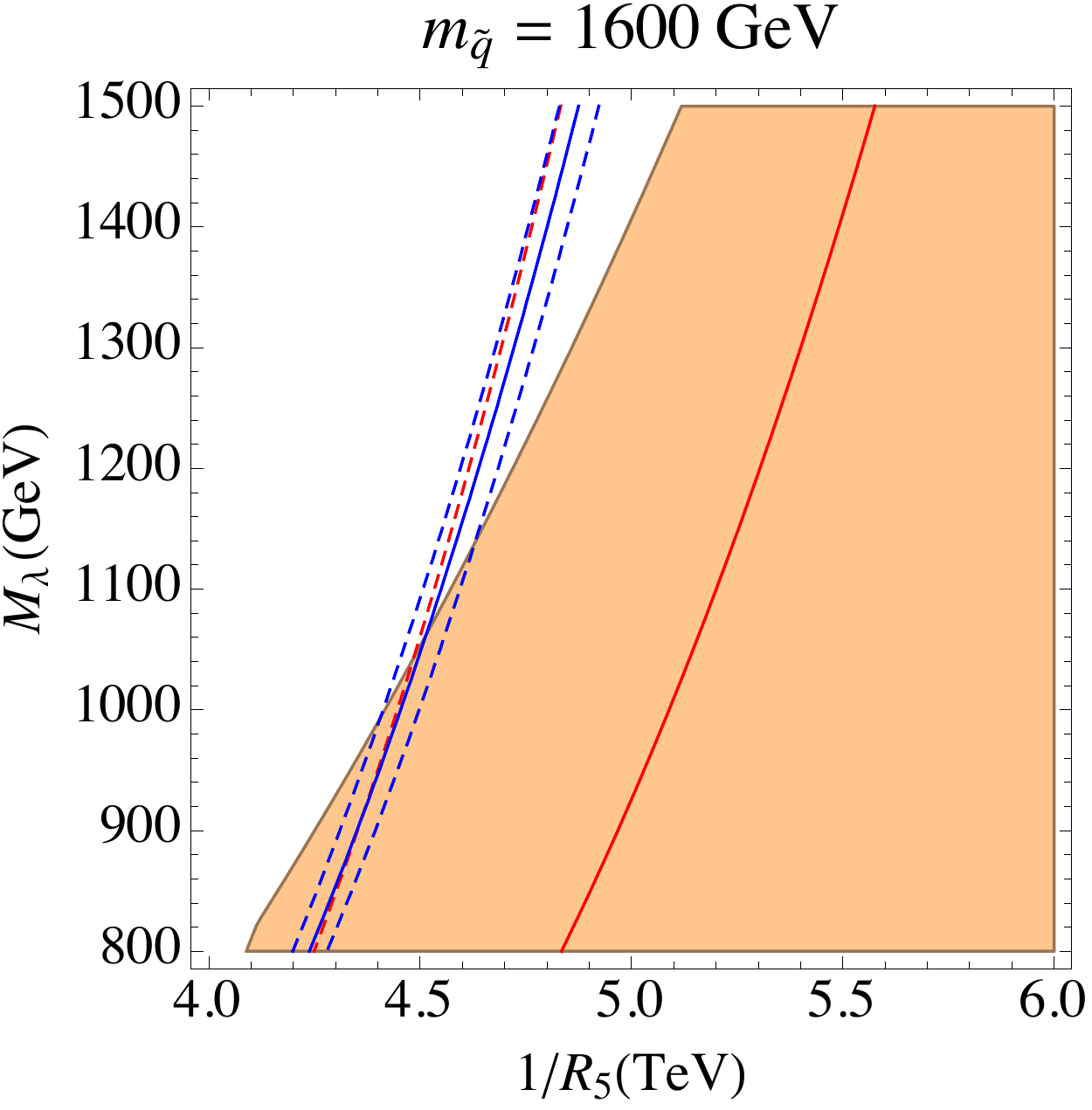}
\includegraphics[width=0.32\columnwidth]{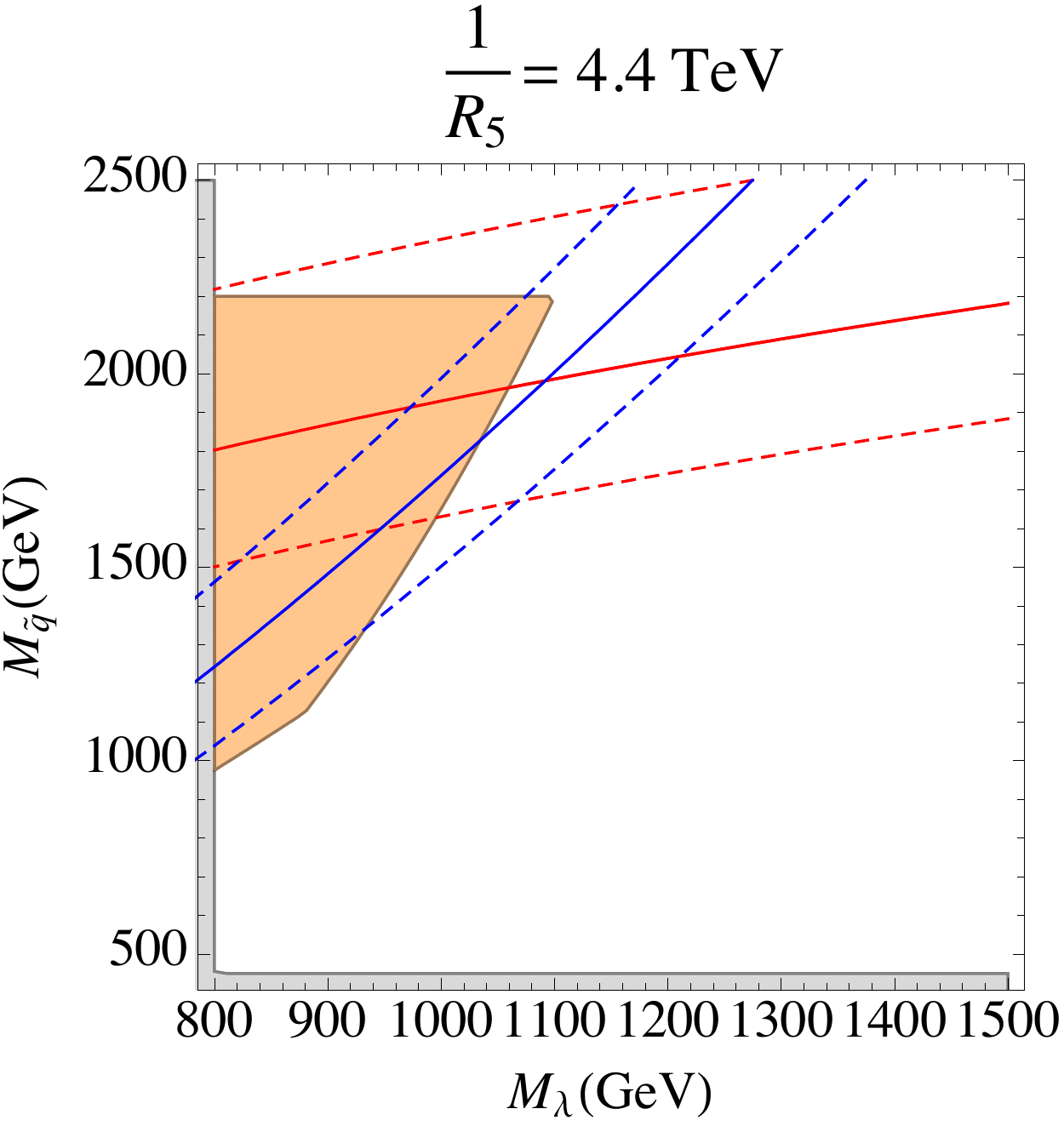}
\includegraphics[width=0.32\columnwidth]{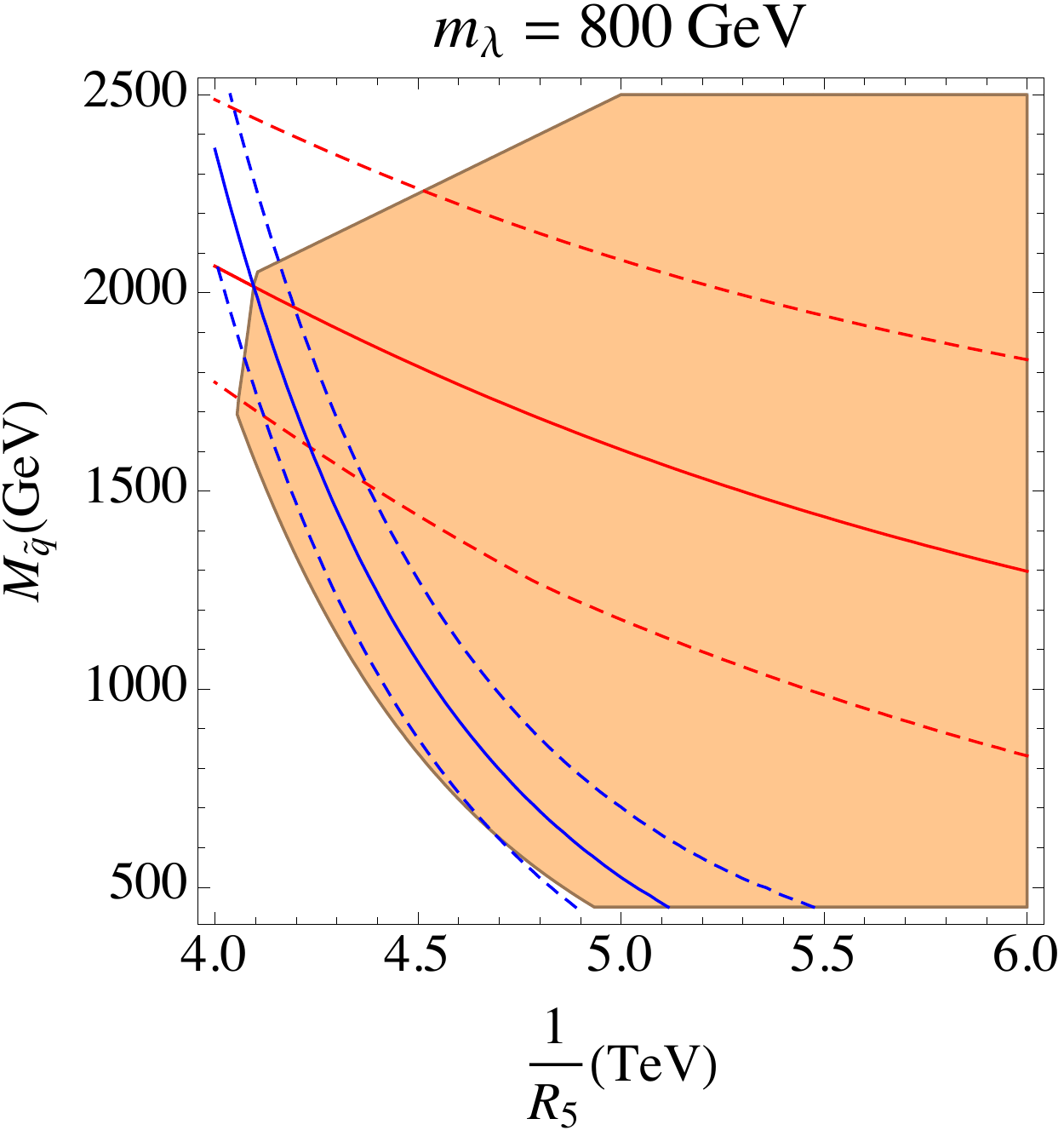}
\caption{These figures show the allowed ranges of the gaugino mass, $m_{\lambda}$, the stop mass, $m_{\tilde{q}}$, and the compactification scale, $1/R_5$, that lead to the experimentally measured physical higgs mass and $\alpha_3$. The light grey regions are the experimentally excluded regions from various collider constraints on the low-lying charged particles, the light orange region is the allowed region that satisfies the unitarity constraint and gives large enough higher dimensional operators to generate the right $\lambda^X$-gaugino mass and stop masses smaller than $1/2R_5$. The red (red-dashed) line marks the central value (contours) of higgs mass that agrees with experiment and the blue (blue-dashed) line marks the central value (contours) of observed $\alpha_s$ taking into account the various theoretical and experimental uncertainties. {\it Left panel:} The relationship between the gaugino mass $m_{\lambda^X}$ and the compactification scale $1/R_5$ fixing $m_{\tilde{q}} =1600\GeV$, the missing red-dashed line is further to the right of the plot. {\it Middle panel:} The relationship between the gaugino mass $m_{\lambda^X}$ and the squark masses $m_{\tilde{q}}$ fixing $1/R_5 = 4.4 \TeV$. {\it Right panel:} The relationship between the squark masses $m_{\tilde{q}}$ and the compactification scale $1/R_5$ fixing the gaugino mass to be $m_{\lambda^X} =800 \GeV$.}\label{Fig::relation}
\end{figure}

\begin{figure}[htb]
\centering
\begin{minipage}[t]{0.4\textwidth}
\includegraphics[width=\textwidth]{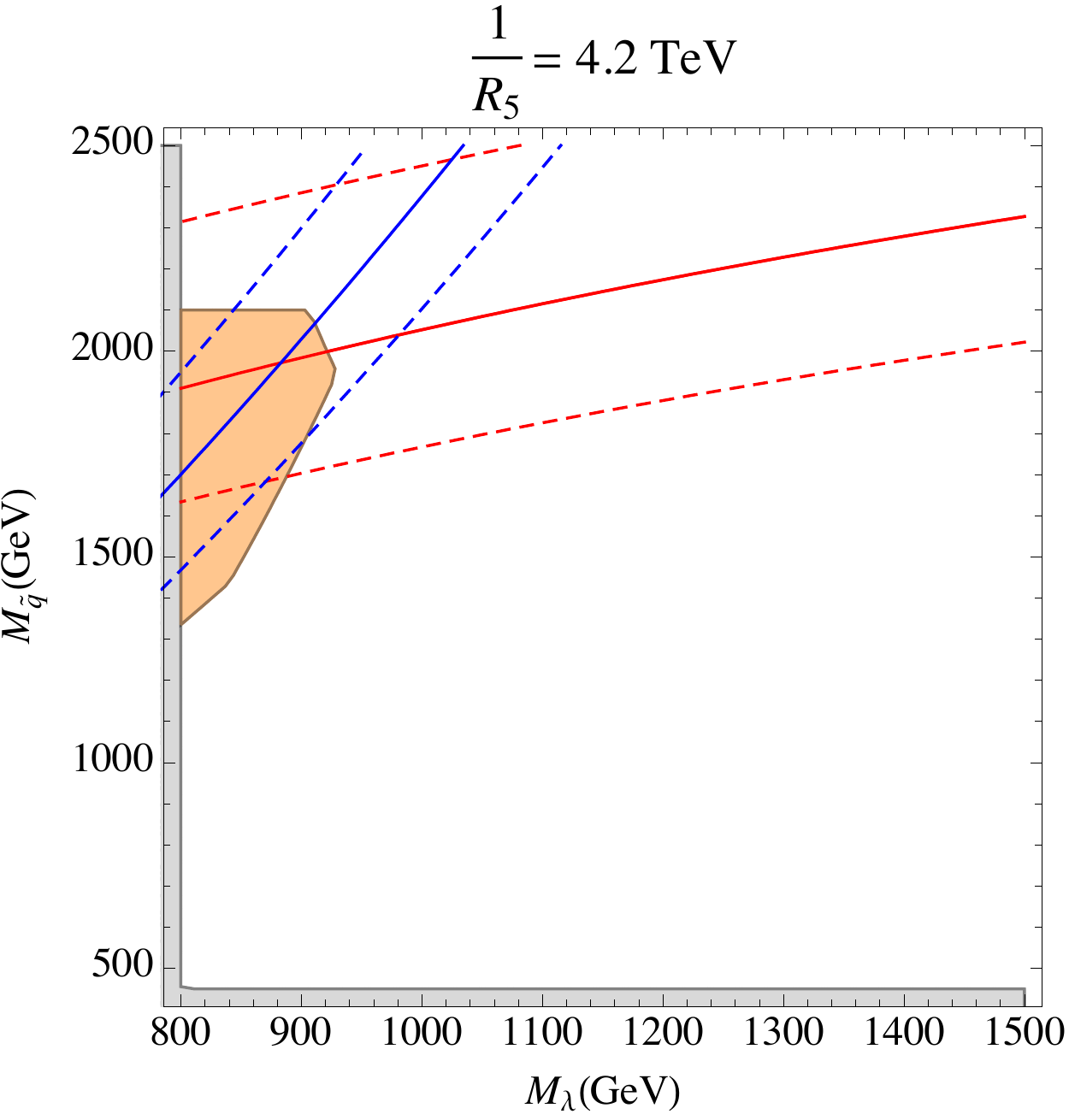}
\end{minipage}
\begin{minipage}[t]{0.4\textwidth}
\centering
\includegraphics[width=\textwidth]{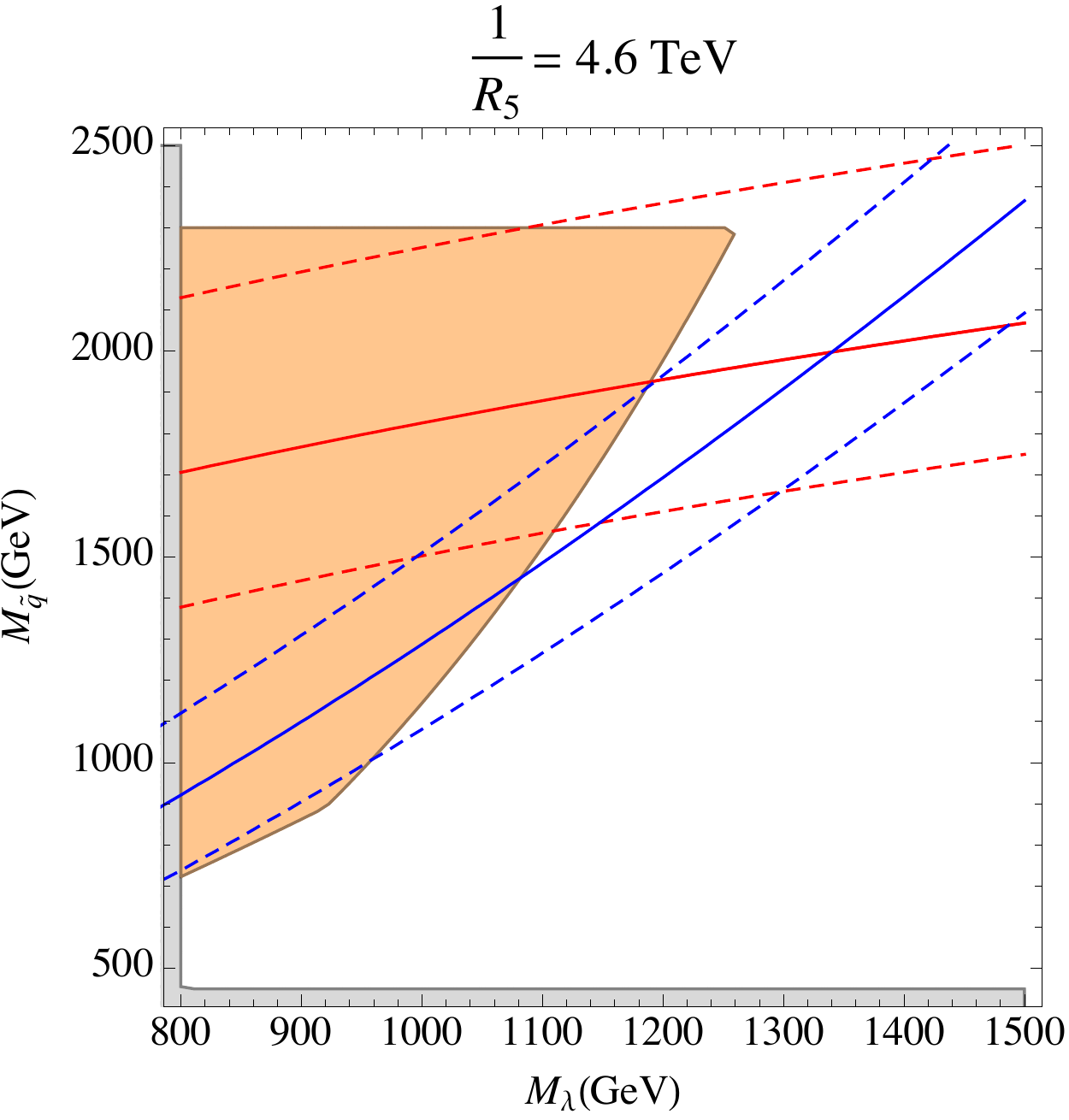}
\end{minipage}
\caption{Similar to Figure~\ref{Fig::relation}. {\it Left panel:} The relationship between the gaugino $\lambda^X$ mass and the squark masses $m_{\tilde{q}}$ taking $1/R_5 = 4.2 \TeV$. {\it Right panel:} The relationship between the gaugino $\lambda^X$ mass and the squark masses $m_{\tilde{q}}$ taking $1/R_5 = 4.6 \TeV$.}\label{Fig::masslimit}
\end{figure}

\subsection{Summary and unification favored particle spectrum}

The unification scenario predicts new particles with distinct collider signatures that can be looked for in the next LHC run. Unification (assuming no additional field content below $M_{\rm GUT}$ that adds to the $SU(6)$ breaking by orbifold bcs and brane localized matter) predicts a relationship between the masses of the light modes that are of experimental interest. In Figure~\ref{Fig::relation}, we demonstrate the mass relationship between the gaugino mass, $m_{\lambda^X}$, the size of the fifth dimension, $1/R_5$, and the squark masses, $m_{\tilde{q}}$, keeping the heavy higgs masses $m_{H_{u,d}}, m_{S_{u,d}}$ fixed at $400 \GeV$\footnote{The effect of the change of the heavy higgs masses is small because of both the low multiplicity of the states and the small range of masses within which they are both experimentally and theoretically allowed}. These plots provide a guide to experimentally search for signatures of a low scale unified model with a natural spectrum.

The unification of gauge couplings and the correct prediction for the higgs mass requires that the compactification scale to be $\frac{1}{R_5} \gtrsim 2.2 \TeV$ so that the zero mode $\lambda^X$-gaugino and the squarks masses required are not experimentally excluded (here taking into account auto-concealment~\cite{Dimopoulos:2014psa}) and also that the $Z'$ mass is not excluded by searches for di-jet resonances at LHC, shown in
Figure~\ref{Fig::masslimit}.  As the compactification scale increase, unification and the higgs mass prefers smaller squark masses and larger gaugino masses and when $1/R_5$ approaches $5.8 \TeV$, the size of the higher dimensional operators in the theory is no longer large enough to generate the zero mode gaugino mass needed to have unification, shown in Figure~\ref{Fig::masslimit}.

To conclude, unification and the requirement to get the higgs mass from non-decoupling D-term of $U(1)'$ sets stringent constraint on the viable parameter space that produces precision unification in six dimensions. We are aware of new vector-like pairs of leptons may also generate one-loop contribute to the physical higgs mass in the MNSUSY setup~\cite{Garcia:2015sfa}. These particles might change the relative running between the $\SUC$ and $\EW$, we will not discuss that case in this paper.


\section{Conclusion and Remarks}

In this paper we have constructed and studied a unified version of the MNSUSY theory.  At the first stage of unification $SU(2)_{\rm L} \times U(1)_{\rm Y}$ is unified into a 5D $N=2$ supersymmetric $SU(3)_{EW}$ gauge theory on an orbifold line segment of size $\pi R_5$.  Despite the higher-dimensional nature
of the theory above the energy scale $1/R_5\sim 4.4\TeV$ (the energy scale at which Scherk-Schwarz supersymmetry breaking also takes place), the differential running between the $SU(2)_{\rm L}$ and $U(1)_{\rm Y}$ gauge couplings is \emph{calculable and logarithmic} and gives the small corrections necessary to move the $SU(3)_{EW}$ prediction of $\sin^2 \theta_W =1/4$ to the experimentally determined LEP value.  Specifically, in Section~2 we show that $\sin^2 \theta_W (M_Z)=0.2312$ is correctly predicted to within $2\%$ uncertainty in the region where electroweak symmetry breaking is better than $\sim 10 \%$ tuned.  We stress that, in the context of MNSUSY (in contrast to some previous uses of $SU(3)_{EW}$~\cite{Dimopoulos:2002jf,Dimopoulos:2002mv,Li:2002pb,Hall:2002rk}) this unified prediction is not `tuned' to reproduce the right value $\sin^2 \theta_W (M_Z)$ by freely varying some continuous parameter of the theory.  In particular, the size, $R_5$, of the 5th dimension is \emph{independently fixed} by the requirement that
Scherk-Schwarz supersymmetry breaking leads to successful electroweak symmetry breaking without excessive fine-tuning.  Moreover,
the parametrically large size of the 5th dimension compared with the fundamental scale of the theory guarantees that any brane localized operators that do not respect the bulk symmetry are volume suppressed, and do not spoil the precision prediction of $\sin^2 \theta_W (M_Z)$.

At the second stage of unification, $SU(3)_{EW}\times SU(3)_C$ (as well as an additional $U(1)'$ necessary for lifting the physical Higgs state to its
observed mass -- see Section~4) unifies into a six-dimensional $N=4$ supersymmetric $SU(6)$ gauge theory defined on a $S^1/(Z_2\times Z_2') \times S^1/(Z_2 \times Z_2')$ 
rectangular orbifold of extra-dimensional size $\pi^2 R_5 R_6$. Both logarithmic running and \emph{linear} threshold corrections
differentially correcting the gauge couplings.  While, as we argue in Section~3, the enlarged gauge symmetry and supersymmetry in the 6D bulk ensure that the both the differential logarithmic running and linear threshold corrections are calculable and various higher dimensional operators that spoil precision unification are not present in the bulk unified theory, the corrections due to $SU(6)$-violating, and especially $Z_2$ violating, brane-localized operators are not so suppressed as in the $SU(3)_{EW}$ case.
(Due to the fact that $1/R_6 \sim 40\TeV$ is substantially closer to the fundamental scale of the underlying UV theory.)  
Correspondingly, we find the derived value of $\alpha_s(M_Z)$ is known to only $\pm 7\%$, reflecting the unusual two-step unification.

As discussed in Section~5, the unification scenario predicts new particles with distinct collider signatures that can be looked for in present and future LHC runs.
The allowed mass range for these light modes and the compactification scale are summarized in Table~\ref{Tab::mass}.  The squark and $\lambda^X$-gaugino masses that lead to precision unification are all within reach of the LHC, producing distinct signatures in their decays.

\begin{table}
\centering
\begin{tabular}{|c|c|c|c|c|c|c|}
\hline 
 ${}$ & $m_{\tilde{q}} (\GeV)$ & $m_{\lambda^X} (\GeV)$ &$m_{S_{u,d}} (\GeV)$ &$m_{Z'} (\TeV)$ & $\frac{1}{R_5} (\TeV)$ & $\frac{1}{R_6} (\TeV)$   \tabularnewline
\hline 
BM & $1600$ & $800$ & $400$ & $2.5$ & $4.4$ & $\sim 40$ \tabularnewline
\hline 
Range & $({\it 1400},{\it 2300})$ & $(800,{\it 1200})$ & $(400, {\it 500})$ & $(2.3,{\it 3.5})$ & $({\it 4.1}, {\it 4.7})$ & ${}$  
\tabularnewline
\hline
\end{tabular}\caption{Benchmark (BM) value of various mass and energy scales in the theory leading to full unification in 6D at $\sim 120 \TeV$ with the correct physical higgs mass, and with better than 10\% tuning of EWSB.  The mass range of the low lying states that satisfies all current experimental (roman numbers) and theoretical and fine-tuning constraints (italic numbers) is also provided.  In addition to requiring low fine-tuning, here we also assume higher dimensional operators can be present with at most NDA value coefficients.  Thus masses of particles exceeding these ranges are possible if unexplained tuning is allowed.}\label{Tab::mass}
\end{table}

Finally, we remark that unification of $\EW$ and $\SUC$ via a differential \emph{linear} threshold effect is a necessity because of the low fundamental scale required by perturbative unitarity considerations. However, following the work of Seiberg~\cite{Seiberg:1996bd}, progress~\cite{Seiberg:1996bd,Intriligator:1997pq,Seiberg:1996qx,Danielsson:1997kt} has been made towards the construction of exact finite theories in both five and six dimensions with non-trivial interacting fixed points. These studies might open up new possibilities for UV completions of MNSUSY into a theory with a non-trivial interacting fixed point, and, in that case, it is possible that the complete unification of MNSUSY occurs via purely logarithmic running, as in the MSSM, though of a highly unusual extra-dimensional form.


\acknowledgments
 We are especially grateful to Kiel Howe for sharing with us many insights on MNSUSY, and to Isabel Garcia Garcia for collaboration at early stages of the project and many important discussions.  We also thank Asimina Arvanitaki, Masha Baryakhtar, Savas Dimopoulos, Xinlu Huang, James Scoville, Ken van Tilburg and Yue Zhao for helpful discussions.  We are both grateful to the CERN theory group, and JMR additionally thanks SITP Stanford, for kind hospitality during portions of this project.  JH particularly thanks Michelle Connor and Nanie Perrin for help with logistics during his stay in CERN. JH thank NSF grant PHYS-1316699 for support.


\appendix


\section{Renormalization scheme and scheme dependent constants}

In this paper, the logarithmic running of the gauge couplings and the crossing of multiple mass thresholds is treated in the $\overline{\rm DR}$ scheme, in which
no finite threshold corrections are present~\cite{Antoniadis:1982vr}.  On the other hand, the gauge couplings are usually given in the $\overline{\rm MS}$ scheme~\cite{Beringer:1900zz}
\begin{align}
{\alpha_{\rm EM}^{\overline{\rm MS}} (M_Z)}^{-1} &= 127.944 \pm 0.014 \nonumber \\
\sin^2 \theta_W^{\overline{\rm MS}}(M_Z) &= 0.23116 \pm 0.00012 \nonumber \\
\alpha_{3}^{\overline{\rm MS}}(M_Z) &= 0.118 \pm 0.003
\end{align}
which translate to the $\EW \times \SUC$ normalization to be
\begin{align}
{\alpha_{1}^{\overline{\rm MS}} (M_Z)}^{-1} &= 32.790 \pm 0.006 \nonumber\\
{\alpha_{2}^{\overline{\rm MS}} (M_Z)}^{-1} &= 29.576 \pm 0.016 \nonumber\\
\alpha_{3}^{\overline{\rm MS}}(M_Z) &= 0.118 \pm 0.003
\end{align}
The matching between the $\overline{\rm MS}$ scheme and $\overline{\rm DR}$ scheme at $M_Z$ is determined purely by the low energy spectrum below $M_Z$.  Specifically (utilizing the results of~\cite{Antoniadis:1982vr,Martin:1993yx,Baer:2002ek})
\begin{align}
{\alpha_{1}^{\overline{\rm DR}} (M_Z)}^{-1} &= {\alpha_{1}^{\overline{\rm MS}} (M_Z)}^{-1} \nonumber\\
{\alpha_{2}^{\overline{\rm DR}} (M_Z)}^{-1} &= {\alpha_{2}^{\overline{\rm MS}} (M_Z)}^{-1} + \frac{1}{6\pi} \nonumber\\
 {\alpha_{3}^{\overline{\rm DR}} (M_Z)}^{-1} &=  {\alpha_{3}^{\overline{\rm MS}} (M_Z)}^{-1} + \frac{3}{12 \pi}\nonumber\\
\end{align}

Finally, the top quark mass in the $\overline{\rm MS}$ scheme is taken to be $160^{+5}_{-4} \GeV$~\cite{Garcia:2015sfa} in the calculation of the physical higgs mass and coupling evolution.


\section{Generating masses for the exotic states}\label{app:exoticmasses}

\begin{table}
\centering
\begin{tabular}{|c|c c c c c|c c c c|c c|c c|}
\hline 
Field  & Q & $ u^c$ & $d^c$ & $L$ & $e^c$ & $H_{u}$ & $H_{d}$ & $H_{u}^c$ & $H_{d}^c$ & X & S & $\phi_+$ & $\phi_-$ \tabularnewline
\hline 
 $U(1)_R$ & 1 & 1 & 1 & 1 & 1 & 0 & 2 & 2 & 0 & 2 & 2 & 0 & 0 \tabularnewline
 $U(1)'$  & 0 & -1 & 1 & 1/2 & 1/2 & 1 & -1 & -1 & 1 & 0 & 0 & $N_{\phi}$ & -$N_{\phi}$ \tabularnewline
\hline 
\end{tabular}\caption{$U(1)_R$ and $U(1)'$ charge assignments of the bulk and brane localized fields to be consistent with the superpotential and ${\rm K\ddot{a}hler}$ potential terms in the paper}\label{Table::R}
\end{table}

Operators on the $z=0$ brane are required to generate Yukawa couplings and squark masses.  In addition two brane-localized SM singlets $X$ and $S$ are needed to raise the masses of the exotic zero modes of the gaugino $\lambda^X$ and singlet higgses ${S}_{u,d}$.  Specifically, in terms of the 4d superpotential and ${\rm K\ddot{a}hler}$ potential on $y = \pi R_5, z=0$ brane:
\begin{align}
W &\supset \int {\rm d} y {\rm d} z \,\delta(y-\pi R_5) \delta(z) \, \frac{y_u}{M_*^{1/2}} {H}_u Q u^c  \nonumber\\
K &\supset \int  {\rm d} y {\rm d} z  \,\delta(y-\pi R_5) \delta(z) \left\{\frac{y_d}{M_*^{3/2}} \left({H}_u^{\dagger} X^{\dagger} Q d^c\right) + \frac{y_L}{M_*^{5/2}} \left({H}_u^{\dagger} X^{\dagger} L e^c \right) \right\}
\label{Eq:yukawa}
\end{align}
where $X$ develops a SUSY breaking F-term at $O(1/R_5)$.  The bulk higgses and brane localized squarks get masses through the higher dimensional operators in the ${\rm K\ddot{a}hler}$ potential on the $y = \pi R_5, z=0$ brane
\begin{align}
K &\supset \int {\rm d} y {\rm d} z \,\delta(y-\pi R_5) \delta(z) \left\{ \frac{1}{M_*^{2}} \left( c_Q X^{\dagger} X Q^{\dagger} Q+ c_u X^{\dagger} X u^{c \dagger} u^c + c_d X^{\dagger} X d^{ c \dagger} d^c \right) \right.\nonumber \\
 &\left. + \frac{1}{M_*^{3}} \left(c_{H_u}X^{\dagger} X H_u^{\dagger} H_u + c_{H_d}X^{\dagger} X H_d^{\dagger} H_d +c_{S_u}X^{\dagger} X S_u^{\dagger} S_u + c_{S_d}X^{\dagger} X S_d^{\dagger} S_d\right) \right\}
\label{Eq:Mscalars}
\end{align}
These operators also provide both tree level and one loop (through the stop loop) contributions to the higgs soft-mass-squared and triggers electroweak symmetry breaking~\cite{Dimopoulos:2014aua,Garcia:2015sfa}.  The $U(1)_R$ symmetry is explicitly broken by higher-dimensional superpotential operators on the $y = 0, z=0$ brane
\begin{align}
W \supset \int {\rm d} y {\rm d} z \delta(y) \delta(z) \frac{c_{\lambda^X}}{M_*}S  W^{\alpha} W_{\alpha},
\label{Eq:Mgaugino}
\end{align}
where $S$ develops a SUSY breaking F-term at less than $1/2R_5$ to generate the required gaugino masses, with $W^{\alpha}$ defined by the gauge kinetic term in 6D,
\begin{align}
W \supset \int {\rm d} y {\rm d} z \frac{1}{g_6^2} W^{\alpha} W_{\alpha}.
\end{align}
There are also other $U(1)_R$ symmetry breaking higher-dimensional operators that generates a $B\mu$-term for the Higgs, and a much suppressed A-term due to locality.
The structure suggests that the mass of the squarks and gauginos are likely comparable  while the higgs ($\mathcal{H}_{u,d}$) are a factor of $\frac{1}{(\pi M_* R_5)^{1/2}}$ smaller. 
The Lagrangian of the theory to reproduce the low energy mass spectrum of the SM matter can be written down on the $\EW$ broken brane as
\begin{align}
\mathcal{L} &= \int {\rm d} y {\rm d} z \delta(y-\pi R_5) \delta (z) \left( y_u H_u Q u^c + y_d H_u^{\dagger} Q d^c + y_L H_u^{\dagger} L e^c + M_{{S}_{u,d}}^2 {S}_{u,d}^{\dagger} {S}_{u,d} + M_{{H}_{u,d}}^2 {H}_{u,d}^{\dagger} {H}_{u,d} \right) \nonumber\\
&+  \int {\rm d} y {\rm d} z \delta(y) \delta (z) M_{\lambda^X} \lambda^X \lambda^X \cdots
\end{align}


\bibliography{unification}

\providecommand{\href}[2]{#2}\begingroup\raggedright\begin{thebibliography}{100}

\bibitem{Dimopoulos:2014aua}
S.~Dimopoulos, K.~Howe, and J.~March-Russell, {\it {Maximally Natural
  Supersymmetry}},  {\em Phys.Rev.Lett.} {\bf 113} (2014) 111802,
  [\href{http://arxiv.org/abs/1404.7554}{{\tt arXiv:1404.7554}}].

\bibitem{Garcia:2014lfa}
I.~Garcia~Garcia and J.~March-Russell, {\it {Rare Flavor Processes in Maximally
  Natural Supersymmetry}},  {\em JHEP} {\bf 1501} (2015) 042,
  [\href{http://arxiv.org/abs/1409.5669}{{\tt arXiv:1409.5669}}].

\bibitem{Garcia:2015sfa}
I.~Garcia~Garcia, K.~Howe, and J.~March-Russell, {\it {Natural Scherk-Schwarz
  Theories of the Weak Scale}},  {\em JHEP} {\bf 12} (2015) 005,
  [\href{http://arxiv.org/abs/1510.07045}{{\tt arXiv:1510.07045}}].

\bibitem{Gherghetta:2012gb}
T.~Gherghetta, B.~von Harling, A.~D. Medina, and M.~A. Schmidt, {\it {The
  Scale-Invariant NMSSM and the 126 GeV Higgs Boson}},  {\em JHEP} {\bf 1302}
  (2013) 032, [\href{http://arxiv.org/abs/1212.5243}{{\tt arXiv:1212.5243}}].

\bibitem{Arvanitaki:2013yja}
A.~Arvanitaki, M.~Baryakhtar, X.~Huang, K.~van Tilburg, and G.~Villadoro, {\it
  {The Last Vestiges of Naturalness}},  {\em JHEP} {\bf 1403} (2014) 022,
  [\href{http://arxiv.org/abs/1309.3568}{{\tt arXiv:1309.3568}}].

\bibitem{Feng:2013pwa}
J.~L. Feng, {\it {Naturalness and the Status of Supersymmetry}},  {\em
  Ann.Rev.Nucl.Part.Sci.} {\bf 63} (2013) 351--382,
  [\href{http://arxiv.org/abs/1302.6587}{{\tt arXiv:1302.6587}}].

\bibitem{Fan:2014txa}
J.~Fan and M.~Reece, {\it {A New Look at Higgs Constraints on Stops}},  {\em
  JHEP} {\bf 1406} (2014) 031, [\href{http://arxiv.org/abs/1401.7671}{{\tt
  arXiv:1401.7671}}].

\bibitem{Gherghetta:2014xea}
T.~Gherghetta, B.~von Harling, A.~D. Medina, and M.~A. Schmidt, {\it {The price
  of being SM-like in SUSY}},  {\em JHEP} {\bf 1404} (2014) 180,
  [\href{http://arxiv.org/abs/1401.8291}{{\tt arXiv:1401.8291}}].

\bibitem{Hardy:2013ywa}
E.~Hardy, {\it {Is Natural SUSY Natural?}},  {\em JHEP} {\bf 1310} (2013) 133,
  [\href{http://arxiv.org/abs/1306.1534}{{\tt arXiv:1306.1534}}].

\bibitem{Scherk:1978ta}
J.~Scherk and J.~H. Schwarz, {\it {Spontaneous Breaking of Supersymmetry
  Through Dimensional Reduction}},  {\em Phys.Lett.} {\bf B82} (1979) 60.

\bibitem{Scherk:1979zr}
J.~Scherk and J.~H. Schwarz, {\it {How to Get Masses from Extra Dimensions}},
  {\em Nucl.Phys.} {\bf B153} (1979) 61--88.

\bibitem{Antoniadis:1998sd}
I.~Antoniadis, S.~Dimopoulos, A.~Pomarol, and M.~Quiros, {\it {Soft masses in
  theories with supersymmetry breaking by TeV compactification}},  {\em
  Nucl.Phys.} {\bf B544} (1999) 503--519,
  [\href{http://arxiv.org/abs/hep-ph/9810410}{{\tt hep-ph/9810410}}].

\bibitem{Delgado:1998qr}
A.~Delgado, A.~Pomarol, and M.~Quiros, {\it {Supersymmetry and electroweak
  breaking from extra dimensions at the TeV scale}},  {\em Phys.Rev.} {\bf D60}
  (1999) 095008, [\href{http://arxiv.org/abs/hep-ph/9812489}{{\tt
  hep-ph/9812489}}].

\bibitem{Pomarol:1998sd}
A.~Pomarol and M.~Quiros, {\it {The Standard model from extra dimensions}},
  {\em Phys.Lett.} {\bf B438} (1998) 255--260,
  [\href{http://arxiv.org/abs/hep-ph/9806263}{{\tt hep-ph/9806263}}].

\bibitem{Delgado:2001si}
A.~Delgado and M.~Quiros, {\it {Supersymmetry and finite radiative electroweak
  breaking from an extra dimension}},  {\em Nucl.Phys.} {\bf B607} (2001)
  99--116, [\href{http://arxiv.org/abs/hep-ph/0103058}{{\tt hep-ph/0103058}}].

\bibitem{Barbieri:2003kn}
R.~Barbieri, G.~Marandella, and M.~Papucci, {\it {The Higgs mass as a function
  of the compactification scale}},  {\em Nucl.Phys.} {\bf B668} (2003)
  273--292, [\href{http://arxiv.org/abs/hep-ph/0305044}{{\tt hep-ph/0305044}}].

\bibitem{Barbieri:2002sw}
R.~Barbieri, L.~J. Hall, G.~Marandella, Y.~Nomura, T.~Okui, et~al., {\it
  {Radiative electroweak symmetry breaking from a quasilocalized top quark}},
  {\em Nucl.Phys.} {\bf B663} (2003) 141--162,
  [\href{http://arxiv.org/abs/hep-ph/0208153}{{\tt hep-ph/0208153}}].

\bibitem{Barbieri:2002uk}
R.~Barbieri, G.~Marandella, and M.~Papucci, {\it {Breaking the electroweak
  symmetry and supersymmetry by a compact extra dimension}},  {\em Phys.Rev.}
  {\bf D66} (2002) 095003, [\href{http://arxiv.org/abs/hep-ph/0205280}{{\tt
  hep-ph/0205280}}].

\bibitem{Barbieri:2000vh}
R.~Barbieri, L.~J. Hall, and Y.~Nomura, {\it {A Constrained standard model from
  a compact extra dimension}},  {\em Phys.Rev.} {\bf D63} (2001) 105007,
  [\href{http://arxiv.org/abs/hep-ph/0011311}{{\tt hep-ph/0011311}}].

\bibitem{Diego:2005mu}
D.~Diego, G.~von Gersdorff, and M.~Quiros, {\it {Supersymmetry and electroweak
  breaking in the interval}},  {\em JHEP} {\bf 0511} (2005) 008,
  [\href{http://arxiv.org/abs/hep-ph/0505244}{{\tt hep-ph/0505244}}].

\bibitem{Diego:2006py}
D.~Diego, G.~von Gersdorff, and M.~Quiros, {\it {The MSSM from Scherk-Schwarz
  supersymmetry breaking}},  {\em Phys.Rev.} {\bf D74} (2006) 055004,
  [\href{http://arxiv.org/abs/hep-ph/0605024}{{\tt hep-ph/0605024}}].

\bibitem{Gersdorff:2007kk}
G.~von Gersdorff, {\it {The MSSM on the Interval}},  {\em Mod.Phys.Lett.} {\bf
  A22} (2007) 385--398, [\href{http://arxiv.org/abs/hep-ph/0701256}{{\tt
  hep-ph/0701256}}].

\bibitem{Bhattacharyya:2012ct}
G.~Bhattacharyya and T.~S. Ray, {\it {Naturally split supersymmetry}},  {\em
  JHEP} {\bf 1205} (2012) 022, [\href{http://arxiv.org/abs/1201.1131}{{\tt
  arXiv:1201.1131}}].

\bibitem{Quiros:2003gg}
M.~Quiros, {\it {New ideas in symmetry breaking}},
  \href{http://arxiv.org/abs/hep-ph/0302189}{{\tt hep-ph/0302189}}.

\bibitem{Delgado:2001xr}
A.~Delgado, G.~von Gersdorff, and M.~Quiros, {\it {Two loop Higgs mass in
  supersymmetric Kaluza-Klein theories}},  {\em Nucl.Phys.} {\bf B613} (2001)
  49--63, [\href{http://arxiv.org/abs/hep-ph/0107233}{{\tt hep-ph/0107233}}].

\bibitem{Antoniadis:1997zg}
I.~Antoniadis, S.~Dimopoulos, and G.~Dvali, {\it {Millimeter range forces in
  superstring theories with weak scale compactification}},  {\em Nucl.Phys.}
  {\bf B516} (1998) 70--82, [\href{http://arxiv.org/abs/hep-ph/9710204}{{\tt
  hep-ph/9710204}}].

\bibitem{Barbieri:2001dm}
R.~Barbieri, L.~J. Hall, and Y.~Nomura, {\it {Models of Scherk-Schwarz symmetry
  breaking in 5-D: Classification and calculability}},  {\em Nucl.Phys.} {\bf
  B624} (2002) 63--80, [\href{http://arxiv.org/abs/hep-th/0107004}{{\tt
  hep-th/0107004}}].

\bibitem{Contino:2001gz}
R.~Contino and L.~Pilo, {\it {A Note on regularization methods in Kaluza-Klein
  theories}},  {\em Phys.Lett.} {\bf B523} (2001) 347--350,
  [\href{http://arxiv.org/abs/hep-ph/0104130}{{\tt hep-ph/0104130}}].

\bibitem{Delgado:2001ex}
A.~Delgado, G.~von Gersdorff, P.~John, and M.~Quiros, {\it {One loop Higgs mass
  finiteness in supersymmetric Kaluza-Klein theories}},  {\em Phys.Lett.} {\bf
  B517} (2001) 445--449, [\href{http://arxiv.org/abs/hep-ph/0104112}{{\tt
  hep-ph/0104112}}].

\bibitem{Hosotani:1983xw}
Y.~Hosotani, {\it {Dynamical Mass Generation by Compact Extra Dimensions}},
  {\em Phys.Lett.} {\bf B126} (1983) 309.

\bibitem{Hosotani:1983vn}
Y.~Hosotani, {\it {Dynamical Gauge Symmetry Breaking as the Casimir Effect}},
  {\em Phys.Lett.} {\bf B129} (1983) 193.

\bibitem{Hosotani:1988bm}
Y.~Hosotani, {\it {Dynamics of Nonintegrable Phases and Gauge Symmetry
  Breaking}},  {\em Annals Phys.} {\bf 190} (1989) 233.

\bibitem{Kawamura:2000ev}
Y.~Kawamura, {\it {Triplet doublet splitting, proton stability and extra
  dimension}},  {\em Prog.Theor.Phys.} {\bf 105} (2001) 999--1006,
  [\href{http://arxiv.org/abs/hep-ph/0012125}{{\tt hep-ph/0012125}}].

\bibitem{Altarelli:2001qj}
G.~Altarelli and F.~Feruglio, {\it {SU(5) grand unification in extra dimensions
  and proton decay}},  {\em Phys.Lett.} {\bf B511} (2001) 257--264,
  [\href{http://arxiv.org/abs/hep-ph/0102301}{{\tt hep-ph/0102301}}].

\bibitem{Hall:2001pg}
L.~J. Hall and Y.~Nomura, {\it {Gauge unification in higher dimensions}},  {\em
  Phys.Rev.} {\bf D64} (2001) 055003,
  [\href{http://arxiv.org/abs/hep-ph/0103125}{{\tt hep-ph/0103125}}].

\bibitem{Hebecker:2001wq}
A.~Hebecker and J.~March-Russell, {\it {A Minimal S**1 / (Z(2) x Z-prime (2))
  orbifold GUT}},  {\em Nucl.Phys.} {\bf B613} (2001) 3--16,
  [\href{http://arxiv.org/abs/hep-ph/0106166}{{\tt hep-ph/0106166}}].

\bibitem{Hebecker:2001jb}
A.~Hebecker and J.~March-Russell, {\it {The structure of GUT breaking by
  orbifolding}},  {\em Nucl.Phys.} {\bf B625} (2002) 128--150,
  [\href{http://arxiv.org/abs/hep-ph/0107039}{{\tt hep-ph/0107039}}].

\bibitem{Hall:2001xb}
L.~J. Hall and Y.~Nomura, {\it {Gauge coupling unification from unified
  theories in higher dimensions}},  {\em Phys.Rev.} {\bf D65} (2002) 125012,
  [\href{http://arxiv.org/abs/hep-ph/0111068}{{\tt hep-ph/0111068}}].

\bibitem{Dimopoulos:1981zb}
S.~Dimopoulos and H.~Georgi, {\it {Softly Broken Supersymmetry and SU(5)}},
  {\em Nucl.Phys.} {\bf B193} (1981) 150.

\bibitem{Dimopoulos:1981yj}
S.~Dimopoulos, S.~Raby, and F.~Wilczek, {\it {Supersymmetry and the Scale of
  Unification}},  {\em Phys.Rev.} {\bf D24} (1981) 1681--1683.

\bibitem{Dimopoulos:2002jf}
S.~Dimopoulos, D.~E. Kaplan, and N.~Weiner, {\it {Electroweak unification into
  a five-dimensional SU(3) at a TeV}},  {\em Phys.Lett.} {\bf B534} (2002)
  124--130, [\href{http://arxiv.org/abs/hep-ph/0202136}{{\tt hep-ph/0202136}}].

\bibitem{Dimopoulos:2002mv}
S.~Dimopoulos and D.~E. Kaplan, {\it {The Weak mixing angle from an SU(3)
  symmetry at a TeV}},  {\em Phys.Lett.} {\bf B531} (2002) 127--134,
  [\href{http://arxiv.org/abs/hep-ph/0201148}{{\tt hep-ph/0201148}}].

\bibitem{Li:2002pb}
T.-j. Li and W.~Liao, {\it {Weak mixing angle and the SU(3)c x SU(3) model on
  M**4 x S**1 / (Z(2) x Z-prime(2))}},  {\em Phys.Lett.} {\bf B545} (2002)
  147--152, [\href{http://arxiv.org/abs/hep-ph/0202090}{{\tt hep-ph/0202090}}].

\bibitem{Hall:2002rk}
L.~J. Hall and Y.~Nomura, {\it {Unification of weak and hypercharge
  interactions at the TeV scale}},  {\em Phys.Lett.} {\bf B532} (2002)
  111--120, [\href{http://arxiv.org/abs/hep-ph/0202107}{{\tt hep-ph/0202107}}].

\bibitem{Hall:2002qw}
L.~J. Hall and Y.~Nomura, {\it {SO(10) and SU(6) unified theories on an
  elongated rectangle}},  {\em Nucl.Phys.} {\bf B703} (2004) 217--235,
  [\href{http://arxiv.org/abs/hep-ph/0207079}{{\tt hep-ph/0207079}}].

\bibitem{Jiang:2002at}
J.~Jiang, T.-j. Li, and W.~Liao, {\it {Low-energy six-dimensional $N=2$
  supersymmetric SU(6) models on $T^{2}$ orbifolds}},  {\em J.Phys.} {\bf G30}
  (2004) 245--268, [\href{http://arxiv.org/abs/hep-ph/0210436}{{\tt
  hep-ph/0210436}}].

\bibitem{Hall:2001zb}
L.~J. Hall, Y.~Nomura, and D.~Tucker-Smith, {\it {Gauge Higgs unification in
  higher dimensions}},  {\em Nucl.Phys.} {\bf B639} (2002) 307--330,
  [\href{http://arxiv.org/abs/hep-ph/0107331}{{\tt hep-ph/0107331}}].

\bibitem{Hartanto:2005jr}
A.~Hartanto and L.~Handoko, {\it {Grand unified theory based on the SU(6)
  symmetry}},  {\em Phys.Rev.} {\bf D71} (2005) 095013,
  [\href{http://arxiv.org/abs/hep-ph/0504280}{{\tt hep-ph/0504280}}].

\bibitem{Contino:2001si}
R.~Contino, L.~Pilo, R.~Rattazzi, and E.~Trincherini, {\it {Running and
  matching from five-dimensions to four-dimensions}},  {\em Nucl.Phys.} {\bf
  B622} (2002) 227--239, [\href{http://arxiv.org/abs/hep-ph/0108102}{{\tt
  hep-ph/0108102}}].

\bibitem{Chacko:1999hg}
Z.~Chacko, M.~A. Luty, and E.~Ponton, {\it {Massive higher dimensional gauge
  fields as messengers of supersymmetry breaking}},  {\em JHEP} {\bf 0007}
  (2000) 036, [\href{http://arxiv.org/abs/hep-ph/9909248}{{\tt
  hep-ph/9909248}}].

\bibitem{Seiberg:1996bd}
N.~Seiberg, {\it {Five-dimensional SUSY field theories, nontrivial fixed points
  and string dynamics}},  {\em Phys.Lett.} {\bf B388} (1996) 753--760,
  [\href{http://arxiv.org/abs/hep-th/9608111}{{\tt hep-th/9608111}}].

\bibitem{Intriligator:1997pq}
K.~A. Intriligator, D.~R. Morrison, and N.~Seiberg, {\it {Five-dimensional
  supersymmetric gauge theories and degenerations of Calabi-Yau spaces}},  {\em
  Nucl.Phys.} {\bf B497} (1997) 56--100,
  [\href{http://arxiv.org/abs/hep-th/9702198}{{\tt hep-th/9702198}}].

\bibitem{Seiberg:1996qx}
N.~Seiberg, {\it {Nontrivial fixed points of the renormalization group in
  six-dimensions}},  {\em Phys.Lett.} {\bf B390} (1997) 169--171,
  [\href{http://arxiv.org/abs/hep-th/9609161}{{\tt hep-th/9609161}}].

\bibitem{Danielsson:1997kt}
U.~H. Danielsson, G.~Ferretti, J.~Kalkkinen, and P.~Stjernberg, {\it {Notes on
  supersymmetric gauge theories in five-dimensions and six-dimensions}},  {\em
  Phys.Lett.} {\bf B405} (1997) 265--270,
  [\href{http://arxiv.org/abs/hep-th/9703098}{{\tt hep-th/9703098}}].

\bibitem{Hebecker:2002vm}
A.~Hebecker and A.~Westphal, {\it {Power - like threshold corrections to gauge
  unification in extra dimensions}},  {\em Annals Phys.} {\bf 305} (2003)
  119--136, [\href{http://arxiv.org/abs/hep-ph/0212175}{{\tt hep-ph/0212175}}].

\bibitem{Hebecker:2004xx}
A.~Hebecker and A.~Westphal, {\it {Gauge unification in extra dimensions: Power
  corrections vs. higher-dimension operators}},  {\em Nucl.Phys.} {\bf B701}
  (2004) 273--298, [\href{http://arxiv.org/abs/hep-th/0407014}{{\tt
  hep-th/0407014}}].

\bibitem{Maloney:2004rc}
A.~Maloney, A.~Pierce, and J.~G. Wacker, {\it {D-terms, unification, and the
  Higgs mass}},  {\em JHEP} {\bf 0606} (2006) 034,
  [\href{http://arxiv.org/abs/hep-ph/0409127}{{\tt hep-ph/0409127}}].

\bibitem{SekharChivukula:2001hz}
R.~S. Chivukula, D.~A. Dicus, and H.-J. He, {\it {Unitarity of compactified
  five-dimensional Yang-Mills theory}},  {\em Phys.Lett.} {\bf B525} (2002)
  175--182, [\href{http://arxiv.org/abs/hep-ph/0111016}{{\tt hep-ph/0111016}}].

\bibitem{Chivukula:2003kq}
R.~S. Chivukula, D.~A. Dicus, H.-J. He, and S.~Nandi, {\it {Unitarity of the
  higher dimensional standard model}},  {\em Phys.Lett.} {\bf B562} (2003)
  109--117, [\href{http://arxiv.org/abs/hep-ph/0302263}{{\tt hep-ph/0302263}}].

\bibitem{Muck:2004br}
A.~Muck, L.~Nilse, A.~Pilaftsis, and R.~Ruckl, {\it {Quantization and high
  energy unitarity of 5-D orbifold theories with brane kinetic terms}},  {\em
  Phys.Rev.} {\bf D71} (2005) 066004,
  [\href{http://arxiv.org/abs/hep-ph/0411258}{{\tt hep-ph/0411258}}].

\bibitem{Puchwein:2003jq}
M.~Puchwein and Z.~Kunszt, {\it {Radiative corrections with 5-D mixed position
  / momentum space propagators}},  {\em Annals Phys.} {\bf 311} (2004)
  288--313, [\href{http://arxiv.org/abs/hep-th/0309069}{{\tt hep-th/0309069}}].

\bibitem{Avramis:2006nb}
S.~D. Avramis, {\it {Anomaly-free supergravities in six dimensions}},
  \href{http://arxiv.org/abs/hep-th/0611133}{{\tt hep-th/0611133}}.

\bibitem{Green:1984bx}
M.~B. Green, J.~H. Schwarz, and P.~C. West, {\it {Anomaly Free Chiral Theories
  in Six-Dimensions}},  {\em Nucl.Phys.} {\bf B254} (1985) 327--348.

\bibitem{Schwarz:1995zw}
J.~H. Schwarz, {\it {Anomaly - free supersymmetric models in six-dimensions}},
  {\em Phys.Lett.} {\bf B371} (1996) 223--230,
  [\href{http://arxiv.org/abs/hep-th/9512053}{{\tt hep-th/9512053}}].

\bibitem{Avramis:2005hc}
S.~D. Avramis and A.~Kehagias, {\it {A Systematic search for anomaly-free
  supergravities in six dimensions}},  {\em JHEP} {\bf 0510} (2005) 052,
  [\href{http://arxiv.org/abs/hep-th/0508172}{{\tt hep-th/0508172}}].

\bibitem{Barbieri:2002ic}
R.~Barbieri, R.~Contino, P.~Creminelli, R.~Rattazzi, and C.~Scrucca, {\it
  {Anomalies, Fayet-Iliopoulos terms and the consistency of orbifold field
  theories}},  {\em Phys.Rev.} {\bf D66} (2002) 024025,
  [\href{http://arxiv.org/abs/hep-th/0203039}{{\tt hep-th/0203039}}].

\bibitem{Marti:2002ar}
D.~Marti and A.~Pomarol, {\it {Fayet-Iliopoulos terms in 5-d theories and their
  phenomenological implications}},  {\em Phys.Rev.} {\bf D66} (2002) 125005,
  [\href{http://arxiv.org/abs/hep-ph/0205034}{{\tt hep-ph/0205034}}].

\bibitem{Barbieri:2001cz}
R.~Barbieri, L.~J. Hall, and Y.~Nomura, {\it {A Constrained standard model:
  Effects of Fayet-Iliopoulos terms}},
  \href{http://arxiv.org/abs/hep-ph/0110102}{{\tt hep-ph/0110102}}.

\bibitem{Preskill:1990fr}
J.~Preskill, {\it {Gauge anomalies in an effective field theory}},  {\em Annals
  Phys.} {\bf 210} (1991) 323--379.

\bibitem{Aad:2015zhl}
{\bf ATLAS, CMS} Collaboration, G.~Aad et~al., {\it {Combined Measurement of
  the Higgs Boson Mass in $pp$ Collisions at $\sqrt{s}=7$ and 8 TeV with the
  ATLAS and CMS Experiments}},  {\em Phys. Rev. Lett.} {\bf 114} (2015) 191803,
  [\href{http://arxiv.org/abs/1503.07589}{{\tt arXiv:1503.07589}}].

\bibitem{Dimopoulos:2014psa}
S.~Dimopoulos, K.~Howe, J.~March-Russell, and J.~Scoville, {\it
  {Auto-Concealment of Supersymmetry in Extra Dimensions}},
  \href{http://arxiv.org/abs/1412.0805}{{\tt arXiv:1412.0805}}.

\bibitem{Aad:2014iza}
{\bf ATLAS} Collaboration, G.~Aad et~al., {\it {Search for supersymmetry in
  events with four or more leptons in $\sqrt{s}$ = 8 TeV pp collisions with the
  ATLAS detector}},  {\em Phys.Rev.} {\bf D90} (2014), no.~5 052001,
  [\href{http://arxiv.org/abs/1405.5086}{{\tt arXiv:1405.5086}}].

\bibitem{Chatrchyan:2014aea}
{\bf CMS} Collaboration, S.~Chatrchyan et~al., {\it {Search for anomalous
  production of events with three or more leptons in $pp$ collisions at
  $\sqrt(s) =$ 8 TeV}},  {\em Phys.Rev.} {\bf D90} (2014), no.~3 032006,
  [\href{http://arxiv.org/abs/1404.5801}{{\tt arXiv:1404.5801}}].

\bibitem{Aad:2014pda}
{\bf ATLAS} Collaboration, G.~Aad et~al., {\it {Search for supersymmetry at
  $\sqrt{s}$=8 TeV in final states with jets and two same-sign leptons or three
  leptons with the ATLAS detector}},  {\em JHEP} {\bf 1406} (2014) 035,
  [\href{http://arxiv.org/abs/1404.2500}{{\tt arXiv:1404.2500}}].

\bibitem{Chatrchyan:2013fea}
{\bf CMS} Collaboration, S.~Chatrchyan et~al., {\it {Search for new physics in
  events with same-sign dileptons and jets in pp collisions at $\sqrt{s}$ = 8
  TeV}},  {\em JHEP} {\bf 1401} (2014) 163,
  [\href{http://arxiv.org/abs/1311.6736}{{\tt arXiv:1311.6736}}].

\bibitem{Beenakker:1999xh}
W.~Beenakker, M.~Klasen, M.~Kramer, T.~Plehn, M.~Spira, et~al., {\it {The
  Production of charginos / neutralinos and sleptons at hadron colliders}},
  {\em Phys.Rev.Lett.} {\bf 83} (1999) 3780--3783,
  [\href{http://arxiv.org/abs/hep-ph/9906298}{{\tt hep-ph/9906298}}].

\bibitem{Choi:1998ut}
S.~Choi, A.~Djouadi, H.~K. Dreiner, J.~Kalinowski, and P.~Zerwas, {\it
  {Chargino pair production in e+ e- collisions}},  {\em Eur.Phys.J.} {\bf C7}
  (1999) 123--134, [\href{http://arxiv.org/abs/hep-ph/9806279}{{\tt
  hep-ph/9806279}}].

\bibitem{Choi:1998ei}
S.~Choi, A.~Djouadi, H.~Song, and P.~Zerwas, {\it {Determining SUSY parameters
  in chargino pair production in e+ e- collisions}},  {\em Eur.Phys.J.} {\bf
  C8} (1999) 669--677, [\href{http://arxiv.org/abs/hep-ph/9812236}{{\tt
  hep-ph/9812236}}].

\bibitem{Kramer:2012bx}
M.~Kramer, A.~Kulesza, R.~van~der Leeuw, M.~Mangano, S.~Padhi, et~al., {\it
  {Supersymmetry production cross sections in $pp$ collisions at $\sqrt{s}=7$
  TeV}},  \href{http://arxiv.org/abs/1206.2892}{{\tt arXiv:1206.2892}}.

\bibitem{Randall:1998uk}
L.~Randall and R.~Sundrum, {\it {Out of this world supersymmetry breaking}},
  {\em Nucl.Phys.} {\bf B557} (1999) 79--118,
  [\href{http://arxiv.org/abs/hep-th/9810155}{{\tt hep-th/9810155}}].

\bibitem{Giudice:1998xp}
G.~F. Giudice, M.~A. Luty, H.~Murayama, and R.~Rattazzi, {\it {Gaugino mass
  without singlets}},  {\em JHEP} {\bf 9812} (1998) 027,
  [\href{http://arxiv.org/abs/hep-ph/9810442}{{\tt hep-ph/9810442}}].

\bibitem{Ibe:2012sx}
M.~Ibe, S.~Matsumoto, and R.~Sato, {\it {Mass Splitting between Charged and
  Neutral Winos at Two-Loop Level}},  {\em Phys.Lett.} {\bf B721} (2013)
  252--260, [\href{http://arxiv.org/abs/1212.5989}{{\tt arXiv:1212.5989}}].

\bibitem{Cirelli:2005uq}
M.~Cirelli, N.~Fornengo, and A.~Strumia, {\it {Minimal dark matter}},  {\em
  Nucl.Phys.} {\bf B753} (2006) 178--194,
  [\href{http://arxiv.org/abs/hep-ph/0512090}{{\tt hep-ph/0512090}}].

\bibitem{ATLAS:2014fka}
{\bf ATLAS Collaboration} Collaboration, G.~Aad et~al., {\it {Searches for
  heavy long-lived charged particles with the ATLAS detector in proton-proton
  collisions at $ \sqrt{s}=8 $ TeV}},  {\em JHEP} {\bf 1501} (2015) 068,
  [\href{http://arxiv.org/abs/1411.6795}{{\tt arXiv:1411.6795}}].

\bibitem{Khachatryan:2015lla}
{\bf CMS} Collaboration, V.~Khachatryan et~al., {\it {Constraints on the pMSSM,
  AMSB model and on other models from the search for long-lived charged
  particles in proton-proton collisions at sqrt(s) = 8 TeV}},
  \href{http://arxiv.org/abs/1502.02522}{{\tt arXiv:1502.02522}}.

\bibitem{Asai:2011wy}
S.~Asai, Y.~Azuma, M.~Endo, K.~Hamaguchi, and S.~Iwamoto, {\it {Stau Kinks at
  the LHC}},  {\em JHEP} {\bf 1112} (2011) 041,
  [\href{http://arxiv.org/abs/1103.1881}{{\tt arXiv:1103.1881}}].

\bibitem{Graham:2012th}
P.~W. Graham, D.~E. Kaplan, S.~Rajendran, and P.~Saraswat, {\it {Displaced
  Supersymmetry}},  {\em JHEP} {\bf 1207} (2012) 149,
  [\href{http://arxiv.org/abs/1204.6038}{{\tt arXiv:1204.6038}}].

\bibitem{CMS:2014gxa}
{\bf CMS} Collaboration, V.~Khachatryan, {\it {Search for disappearing tracks
  in proton-proton collisions at $ \sqrt{s}=8 $ TeV}},  {\em JHEP} {\bf 1501}
  (2015) 096, [\href{http://arxiv.org/abs/1411.6006}{{\tt arXiv:1411.6006}}].

\bibitem{Khachatryan:2014mea}
{\bf CMS} Collaboration, V.~Khachatryan et~al., {\it {Search for Displaced
  Supersymmetry in events with an electron and a muon with large impact
  parameters}},  {\em Phys.Rev.Lett.} {\bf 114} (2015), no.~6 061801,
  [\href{http://arxiv.org/abs/1409.4789}{{\tt arXiv:1409.4789}}].

\bibitem{Aad:2013yna}
{\bf ATLAS} Collaboration, G.~Aad et~al., {\it {Search for charginos nearly
  mass degenerate with the lightest neutralino based on a disappearing-track
  signature in pp collisions at $\sqrt(s)$=8  TeV with the ATLAS
  detector}},  {\em Phys.Rev.} {\bf D88} (2013), no.~11 112006,
  [\href{http://arxiv.org/abs/1310.3675}{{\tt arXiv:1310.3675}}].

\bibitem{Khachatryan:2014fba}
{\bf CMS} Collaboration, V.~Khachatryan et~al., {\it {Search for physics beyond
  the standard model in dilepton mass spectra in proton-proton collisions at
  $\sqrt{s}$ = 8 TeV}},  \href{http://arxiv.org/abs/1412.6302}{{\tt
  arXiv:1412.6302}}.

\bibitem{Aad:2014cka}
{\bf ATLAS} Collaboration, G.~Aad et~al., {\it {Search for high-mass dilepton
  resonances in pp collisions at $\sqrt{s}=8$  TeV with the ATLAS
  detector}},  {\em Phys.Rev.} {\bf D90} (2014), no.~5 052005,
  [\href{http://arxiv.org/abs/1405.4123}{{\tt arXiv:1405.4123}}].

\bibitem{Khachatryan:2015sja}
{\bf CMS} Collaboration, V.~Khachatryan et~al., {\it {Search for resonances and
  quantum black holes using dijet mass spectra in proton-proton collisions at
  $\sqrt{s} =$ 8 TeV}},  {\em Phys.Rev.} {\bf D91} (2015), no.~5 052009,
  [\href{http://arxiv.org/abs/1501.04198}{{\tt arXiv:1501.04198}}].

\bibitem{Aad:2014aqa}
{\bf ATLAS} Collaboration, G.~Aad et~al., {\it {Search for new phenomena in the
  dijet mass distribution using $p-p$ collision data at $\sqrt{s}=8$ TeV with
  the ATLAS detector}},  {\em Phys.Rev.} {\bf D91} (2015), no.~5 052007,
  [\href{http://arxiv.org/abs/1407.1376}{{\tt arXiv:1407.1376}}].

\bibitem{Georgi:1974sy}
H.~Georgi and S.~Glashow, {\it {Unity of All Elementary Particle Forces}},
  {\em Phys.Rev.Lett.} {\bf 32} (1974) 438--441.

\bibitem{ArkaniHamed:1999dc}
N.~Arkani-Hamed and M.~Schmaltz, {\it {Hierarchies without symmetries from
  extra dimensions}},  {\em Phys.Rev.} {\bf D61} (2000) 033005,
  [\href{http://arxiv.org/abs/hep-ph/9903417}{{\tt hep-ph/9903417}}].

\bibitem{Mohapatra:2009wp}
R.~Mohapatra, {\it {Neutron-Anti-Neutron Oscillation: Theory and
  Phenomenology}},  {\em J.Phys.} {\bf G36} (2009) 104006,
  [\href{http://arxiv.org/abs/0902.0834}{{\tt arXiv:0902.0834}}].

\bibitem{Babu:2013yww}
K.~Babu, S.~Banerjee, D.~Baxter, Z.~Berezhiani, M.~Bergevin, et~al., {\it
  {Neutron-Antineutron Oscillations: A Snowmass 2013 White Paper}},
  \href{http://arxiv.org/abs/1310.8593}{{\tt arXiv:1310.8593}}.

\bibitem{BaldoCeolin:1994jz}
M.~Baldo-Ceolin, P.~Benetti, T.~Bitter, F.~Bobisut, E.~Calligarich, et~al.,
  {\it {A New experimental limit on neutron - anti-neutron oscillations}},
  {\em Z.Phys.} {\bf C63} (1994) 409--416.

\bibitem{Takita:1986zm}
{\bf Kamiokande} Collaboration, M.~Takita et~al., {\it {A Search for Neutron -
  Anti-neutron Oscillation in a $^{16}$O Nucleus}},  {\em Phys.Rev.} {\bf D34}
  (1986) 902.

\bibitem{Chung:2002fx}
J.~Chung, W.~Allison, G.~Alner, D.~Ayres, W.~Barrett, et~al., {\it {Search for
  neutron anti-neutron oscillations using multiprong events in Soudan 2}},
  {\em Phys.Rev.} {\bf D66} (2002) 032004,
  [\href{http://arxiv.org/abs/hep-ex/0205093}{{\tt hep-ex/0205093}}].

\bibitem{Kopeliovich:2011aa}
V.~Kopeliovich and I.~Potashnikova, {\it {Restriction on the
  Neutron-Antineutron Oscillations from the SNO Data on the Deuteron
  Stability}},  {\em JETP Lett.} {\bf 95} (2012) 1--5,
  [\href{http://arxiv.org/abs/1112.3549}{{\tt arXiv:1112.3549}}].

\bibitem{Abe:2011ky}
{\bf Super-Kamiokande} Collaboration, K.~Abe et~al., {\it {The Search for
  $n-\bar{n}$ oscillation in Super-Kamiokande I}},
  \href{http://arxiv.org/abs/1109.4227}{{\tt arXiv:1109.4227}}.

\bibitem{Antoniadis:1982vr}
I.~Antoniadis, C.~Kounnas, and K.~Tamvakis, {\it {Simple Treatment of Threshold
  Effects}},  {\em Phys.Lett.} {\bf B119} (1982) 377--380.

\bibitem{Beringer:1900zz}
{\bf Particle Data Group} Collaboration, J.~Beringer et~al., {\it {Review of
  Particle Physics (RPP)}},  {\em Phys.Rev.} {\bf D86} (2012) 010001.

\bibitem{Martin:1993yx}
S.~P. Martin and M.~T. Vaughn, {\it {Regularization dependence of running
  couplings in softly broken supersymmetry}},  {\em Phys.Lett.} {\bf B318}
  (1993) 331--337, [\href{http://arxiv.org/abs/hep-ph/9308222}{{\tt
  hep-ph/9308222}}].

\bibitem{Baer:2002ek}
H.~Baer, J.~Ferrandis, K.~Melnikov, and X.~Tata, {\it {Relating bottom quark
  mass in DR-BAR and MS-BAR regularization schemes}},  {\em Phys.Rev.} {\bf
  D66} (2002) 074007, [\href{http://arxiv.org/abs/hep-ph/0207126}{{\tt
  hep-ph/0207126}}].

\end{thebibliography}\endgroup
\bibliographystyle{JHEP}

\end{document}